# Realizing Fast, Scalable and Reliable Scientific Computations in Grid Environments

*Yong Zhao[1], Ioan Raicu[2], Ian Foster[2,3,4], Mihael Hategan[3], Veronika Nefedova[4], Mike Wilde[3,4]*
*[1]Microsoft Corporation, Redmond, WA, USA*
*[2]Department of Computer Science, University of Chicago, Chicago, IL, USA*
*[3]Computation Institute, University of Chicago, Chicago, IL, USA*
*[4]Mathematics and Computer Science Division, Argonne National Laboratory, Argonne, IL, USA*
*yozha@microsoft.com, iraicu@cs.uchicago.edu, {foster,hategan,nefedova,wilde}@mcs.anl.gov*

## Abstract

The practical realization of managing and executing large scale scientific computations efficiently and reliably is quite challenging. Scientific computations often involve thousands or even millions of tasks operating on large quantities of data, such data are often diversely structured and stored in heterogeneous physical formats, and scientists must specify and run such computations over extended periods on collections of compute, storage and network resources that are heterogeneous, distributed and may change constantly.

We present the integration of several advanced systems: Swift, Karajan, and Falkon, to address the challenges in running various large scale scientific applications in Grid environments. Swift is a parallel programming tool for rapid and reliable specification, execution, and management of large-scale science and engineering workflows. Swift consists of a simple scripting language called SwiftScript and a powerful runtime system that is based on the CoG Karajan workflow engine and integrates the Falkon light-weight task execution service. SwiftScript supports concise specifications of complex parallel computations based on dataset typing and iterations, and dynamic dataset mappings for accessing large scale datasets represented in diverse data formats. Karajan dispatches large scale computations onto different computing resources, where the Falkon service optimizes task throughput and resource efficiency via multi-level scheduling and a streamlined dispatcher that delivers performance not provided by any other system. Microbenchmarks show that Falkon throughput (487 tasks/sec) and scalability (54,000 executors and 1.5 million tasks queued) are one to two orders of magnitude better than other systems used in production Grids.

We showcase the scalability, performance and reliability of the integrated system using application examples drawn from astronomy, cognitive neuroscience and molecular dynamics, which all comprise large number of fine-grained jobs. We show that Swift is able to represent dynamic workflows whose structures can only be determined during runtime and reduce largely the code size of various workflow representations using SwiftScript; schedule the execution of hundreds of thousands of parallel computations via the Karajan engine; and achieve up to 90% reduction in execution time than traditional batch schedulers (e.g. PBS) and comparable performance to MPI execution by dispatching jobs to the Falkon execution service.

# 1. INTRODUCTION

With the advances in scientific instrumentation and simulation, scientific data are growing fast in both data size and data analysis complexity. For instance, the Sloan Digital Sky Survey Data Release 6 [33] includes information about 320 million celestial objects, with 9 terabytes of image data and 9 terabytes of catalog data; whereas the CMS detector being built to run at CERN's Large Hadron Collider is expected to generate over a petabyte of data per year [2]. Data volumes are also increasing dramatically in biology, earth science, medicine, and many other disciplines. A typical scientific analysis needs to operate on either a large collection of data, or a wide variation of parameters. With such scale and complexity, scientific computations can no long afford to be carried out on a single desktop; instead, parallel and distributed computing facilities are being adopted, and cluster and Grid computing have been actively put into practice in science communities.

However, it is not trivial, and sometimes often daunting to realize large scale scientific computations in distributed and parallelized fashion. Consider this painful but familiar scenario: A neuroscientist needs to analyze ten thousand functional magnetic resonance imaging (fMRI) files. The analysis program is a complex PERL script. Files are stored in a collection of UNIX directories, with metadata coded in directory and file names. Local computing facilities are inadequate, as the analysis of just a single file can take many minutes and thus the analysis of the entire dataset would require weeks on a single computer. Thus, the scientist must manually extract files, copy them to a remote cluster, start a home-grown script to dispatch tasks, and check exit codes and output files to see which tasks succeeded and failed. When the computation is completed, the problem of documenting what was done remains. These steps can take days to perform manually, and are error prone.

Our investigation of a few fMRI applications indicate that such ad hoc scripts often require hundreds or even thousands of lines of code (see Section Code Size Reduction) to deal with file format issues, the coordination of data transfers and computing resources, and the tracking of execution processes and exceptions. The result of this code and execution complexity is that many interesting computations involving larger numbers of datasets and/or parameters can not be performed. This example is trivial in complexity when compared with other large scale data analysis collaborations. However, it already identifies many of the challenges that we seek to address in this work:

a) Data are often diversely structured and stored in heterogeneous physical formats (most commonly in collections of files). It is typically not feasible to require that all datasets be converted to a common format or moved to a single repository.

b) Scientific datasets are often operated on by procedures developed under different platforms and environments, using different programming languages. A procedure can be an executable for a particular architecture; a script passed to an interpreter, such as Awk, PERL, Python or a command shell; or a Web service with interface defined by WSDL.

c) The large number of computations also requires tasks to be dispatched and executed on parallel and distributed resources for improved throughput and performance. The scheduling of such data analysis procedures often require extended periods of collective usage of compute, storage, and network resources, which may change over time in availability and capacity. Since manual bookkeeping of the process and tracking of exceptions would be next to impossible, we need automated mechanisms to manage the whole process efficiently and reliably.

d) A set of computations is often just a first step in a long analysis process, especially when multiple studies are performed and/or results are shared with colleagues. It is vital to be able to document data provenance in terms of the data from which it was derived and the computational steps followed to derive it.

Such difficulties motivated us to develop Swift, a parallel programming system that supports rapid and reliable specification, execution and management of large scale scientific computations. A few lines of SwiftScript describe how analysis procedures should be applied to the many fMRI images. Swift tools dispatch the thousands of tasks to TeraGrid [34] nodes. The computations are performed quickly and reliably, and the results are documented with descriptions about how they have been computed.

Swift builds on and includes technology previously distributed as the GriPhyN Virtual Data System [12]. The open source Swift software combines:

- A data model and type system based on the XML dataset typing and mapping model (XDTM) [22] to separate logical data structures from their physical representations.
- A simple scripting language SwiftScript to enable the concise, high-level specifications of complex parallel computations on logically typed datasets, and various mappers for accessing such datasets represented in diverse data formats.
- An execution engine based on CoG Karajan [18] that can manage the dispatch of hundreds of thousands of tasks to thousands of processors, whether on parallel computers, campus grids, or multi-site grids, and optimizes the whole process using adaptive scheduling techniques. Its abstract provider interface also enables Swift to submit jobs to different execution services such as GRAM, batched schedulers (PBS, Condor [35], etc.) and the Falkon provisioning and task execution service [30].

SwiftScript supports dynamic features such as dataset typing and iteration, dynamic dataset mapping and workflow expansion, conditional execution etc, where the Swift system adopts advanced scheduling techniques such as pipelining, clustering, and load balancing. We also integrate Swift with the lightweight Falkon execution service as it enhances Swift with improved task throughput and resource efficiency via multi-level scheduling and a streamlined dispatcher.

We showcase the scalability, performance and reliability of the integrated Swift system using an fMRI analysis workflow from cognitive neuroscience, a photorealistic montage application from the national virtual observatory project [21], and a molecular dynamics application from the biochemistry discipline. We show that swift is able to represent scientific datasets and workflows in simple compact form, and describe the dynamic workflow structure present in the montage workflow, which can be only determined at runtime depending on the output of a certain stage. The resulting code size can be one order of magnitude smaller as compared with ad hoc scripts and MPI coding. By using Falkon as the execution service, Swift also achieves comparable performance to MPI execution in scheduling and executing the montage workflow, and up to 90% reduction in execution time compared with execution over GRAM plus PBS.

# 2. RELATED WORK

Parallel programming models are software technologies that express and execute parallel computations. Popular parallel programming models include shared memory with threads, messaging passing, and data parallel models.

Pthreads [24] is a standardized API for creating and manipulating threads on UNIX-style POSIX systems, but it only has a C language binding, and requires explicit parallelism and significant attention to details. OpenMP [26]is a set of APIs jointly defined and endorsed by many computer hardware and software vendors to support multi-platform shared memory multiprocessing programming in C/C++ and FORTRAN on many architectures including UNIX and Microsoft Windows platforms. It comprises a set of compiler directives, library routines and environmental variables that control program execution. OpenMP is simple and easy to use, but it has scalability issue due to memory architecture limitations, and it lacks reliable error handling.

The Message Passing Interface (MPI) API [23] is the most commonly used message passing model in which a set of tasks use their own local memory during computation and communicate by sending and receiving messages. Calls to MPI functions are embedded in source code; parallelism is specified explicitly by programmers. MPI has been widely used with many conventional programming languages.

In a data parallel model, a set of tasks work collectively on a common data structure, where each task performs the same operation on a different partition of the common dataset. High Performance FORTRAN (HPF) [14] is an example in this category that extends FORTRAN-90 with parallel support such as directives for data distribution, assertions and data parallel constructs.

There are advantages to library and language extension approaches. Programs can be written in familiar languages such as C, FORTRAN, or even scripting languages such as Python that have flexible and powerful constructs and functions; libraries and APIs offer extensive parallel computing support for synchronization, communication, data/task partition etc. However, what we try to achieve in Swift is to provide a simple concise notation that allows easy parallelization and supports the composition of large numbers of parallel computations. We do not need all the constructs and features in a full-fledged conventional language, and we are reluctant to expose the directives for explicit parallelism specification, as they require expertise and attention to the details of parallel programming, which may be difficult for end users. Many of the models also require advanced compilation techniques, which may require significant work when we try to apply them in distributed heterogeneous environments, in which we encounter complex issues of data transfer, scheduling, load balancing, monitoring, provenance tracking and exception handling. All these issues can also make the verification, compilation, debugging and optimization of parallel programs difficult.

These considerations led us to choose to implement Swift as a simple, but new, parallel programming notation, designed to coordinate the activities of executable programs. Our decision to implement a new language has the downside that users will have to learn new language syntax and understand the semantics of language constructs and execution. However, since we define only a limited set of language constructs, such as foreach (used, for example, to specify concurrent application of a function to all members of an array) and procedure composition, the learning curve should not be steep. Also, Swift's implicit parallelism does not require intimate understanding of explicit parallel programming, although we may lose some fine control over data/task partitioning in exchange for conciseness.

MapReduce [8] also provides a programming model and runtime system for the processing of large datasets. It is based on a simple model with just two key functions: "map" and "reduce," borrowed from functional languages. The map function applies a specific operation to each of a set of items, producing a new set of items; a reduce function performs aggregation on a set of items. The MapReduce runtime system automatically partitions input data and schedules the execution of programs in a large cluster of commodity machines. The system is made fault tolerant by checking worker nodes periodically and reassigning failed jobs to other worker nodes. Sawzall [27] is an interpreted language that builds on MapReduce and separates the filtering and aggregation phases for more concise program specification and better parallelization.

Swift and MapReduce/Sawzall share the same goals of providing a programming tool for the specification and execution of large parallel computations on large quantities of data, and facilitating the utilization of large distributed resources. However, the two also differ in several aspects:

*Programming model:* MapReduce only supports key-value pairs as input or output datasets and two types of computation functions – map and reduce. In contrast, Swift provides a type system and allows for the definition of complex data structures and arbitrary computational procedures. Furthermore, because Swift is a language, not a library, it can exploit compile-time transformations to (for example) enable pipelining between different "map" operations.

*Data format:* in MapReduce, input and output data are usually key-value pairs, but it is also possible to define new data sources. Swift provides a more flexible mapping mechanism to map between logical data structures and various physical representations.

*Dataset partition:* Swift does not automatically partition input datasets. Instead, datasets can be organized in structures, and individual items in a dataset can be transferred accordingly along with computations.

*Execution environment:* MapReduce schedules computations within a cluster with a shared Google File System. In contrast, Swift schedules across distributed Grid sites that may span multiple administrative domains, and deals with security and resource usage policy issues.

MapReduce is useful in document processing problems, such as distributed indexing, sorting, and clustering etc, where Swift can be used to coordinate distributed workflows involving different applications.

Our work is also closely related to workflow languages and systems. A distinction is sometimes made between scientific workflows and business workflows, where the former is more concerned with the throughput of data through various stages of programs and applications, and the latter correct, timely and secure execution of business logic.

Examples of scientific workflow languages and systems include Taverna [25], Kepler [19], and Triana [5]. XPDL [38], BPEL [1], and BPML[4] are representatives of a family of process definition languages for describing enterprise business processes and services. DAGMan [6] and Pegasus [9] are two systems that are commonly referred to as workflow systems and that have been widely applied in Grid environments. Neither is concerned with specifying data flow, but both involve data-driven specification of task coordination.

BPEL stands for Business Process Execution Language for Web Services. A BPEL process describes interactions between the process and its partners, and the coordination of the interactions to achieve a business goal. BPEL's data model is based on WSDL messages and XML schema, and it uses XPath for data manipulation. BPEL is starting to be tested in scientific contexts. While BPEL can transfer data as XML messages, for large datasets, data exchange must be handled via separate mechanisms. The BPEL 1.0 specification does not support dataset iterations. Thus, according to Emmerich et al [10], an application with repetitive patterns on a collection of datasets resulted in a BPEL document of 200MB in size, and as a consequence, BPEL is cumbersome if not impossible to write for computational scientists. Although BPEL can use XML Schema to describe data types, it does not provide support for mapping between a logical XML view and arbitrary physical representations.

DAGMan provides a workflow engine that manages Condor jobs organized as directed acyclic graphs (DAGs) in which each edge corresponds to an explicit task precedence. DAGMan is known for its stability; it makes persistent workflow execution status and can survive machine crashes. However it has no knowledge of data flow, and in distributed environment works best with a higher-level, data-cognizant layer. It is based on static workflow graphs and lacks dynamic features such as iteration or conditional execution, although these features are being explored.

Pegasus is primarily a set of DAG transformers. Pegasus planners translate a workflow graph into a location specific DAGMan input file, adding stages for data staging, inter-site transfer and data registration. They can prune tasks for files that already exist, select sites for jobs, and cluster jobs based on various criteria. Pegasus performs graph transformation with the knowledge of the whole workflow graph, where in Swift the structure of a workflow is constructed and expanded dynamically. A static workflow graph allows optimizations based on the global state of the workflow, whereas a dynamic workflow structure allows adaptations to changing environments.

Many workflow languages allow sequential, parallel, and recursive patterns, but do not directly support iteration. Taverna relies on its workflow engine to run a process multiple times when a collection is passed to a singleton-argument process. Kepler uses a 'map' operator to apply a function that operates on singletons to collections. Swift's typing supports flexible iteration over datasets—and also type checking, composition, and selection. The ability to map logical types from/to physical representations is not provided by these languages and systems.

# 3. THE SWIFT SYSTEM

The Swift system is a scalable environment for efficient specification, scheduling, monitoring and tracking of parallel programs. The system comprises a simple scripting language called SwiftScript [40] for concise high level specification of complex parallel computations, and an efficient execution engine built on CoG Karajan [18]. Swift grows out of the GriPhyN Virtual Data System [12], which is an open source Grid workflow system that has been widely applied in many scientific domains. In the following sections, we present the motivation, design principles and technical details about SwiftScript and the Swift runtime system.

## 3.1. SwiftScript

Scientific computations often involve large number of parallel tasks operating on large number of datasets. For instance, a spatial reorientation process on a complete fMRI study could involve 9,000 operations on the images in that study. Such large number of computations are common in scientific practice, for instance to study sensitivity to parameter values (in parameter studies) and/or initial conditions (in ensemble simulations). Users working on such problems frequently struggle with the description of the tasks and datasets, and bookkeeping the computation processes and results.

Our goal in designing SwiftScript is to enable concise, easily parallelizable specification of large scale scientific computations on structured datasets, to simplify the description, maintenance and debugging of parallel program specifications. The design addresses primarily three concerns:

**(1) Dataset structures as first-class objects**

Scientific applications deal primarily with file system based datasets. Analysis procedures read file system based data and produce file system based data. However, we try to avoid encoding complex physical layout and storage formats and the access methods for such data into our SwiftScript, as they complicate the description of application logic, and make programs hard to understand and maintain. To this end, we have developed the XML Dataset Typing and Mapping model that allows logical datasets to be described and manipulated as first-class objects in the language. XDTM provides the abstraction from physical representations and locations, and also supports typing of logical datasets. SwiftScript language constructs allow programmers to define variables that correspond to file system structures (XDTM supports other data structures such as relational database schemas as well) and then to operate on those variables as if they were ordinary language objects.

**(2) Interfaces to diverse computational procedures that are rich enough to allow remote execution**

Scientific applications involve many kinds of computational procedures that are developed in different programming languages on different platforms. SwiftScript defines constructs that allow programmers to define interfaces to such procedures and then invoke them from SwiftScript programs. Procedures are modeled as functional (in the sense that they do not modify their inputs and the computations are deterministic), they are defined as logical operations on typed logical datasets. Although this choice limits the application of SwiftScript in non-functional scenarios, it greatly simplifies the programming model and implementation. The interfaces to invoking actual computational procedures provide support for constructing the command-line calls that may be used to execute procedures and for specifying what file system objects may be required at remote execution sites, so that they can be translated into data movements.

**(3) Coordination constructs that allow for implicit parallelism, via functional operations, dataflow, iterations etc.**

Functional procedures can be composed to form more complex analysis workflows. SwiftScript is a data-flow language in that procedures produce datasets that can be further consumed by other procedures, thus forming data dependencies. The structural representation of a dataset also allows parallel iterations over a collection of data items. SwiftScript does not provide explicit parallel directives but relies on a compiler to exploit parallelism, in relieving programmers from synchronization, communication and task partition issues.

We illustrate in Figure 1 the features and language constructs of SwiftScript using excerpts from an fMRI workflow.

## 3.2. Dataset Typing

Lines 1-6 in Figure 1 declare the logical data types used in the fMRI workflow. An fMRI *Run* is a series of brain scans called volumes, with a *Volume* containing a 3D image of a volumetric slice of a brain image, represented by an *Image* (voxels) and a *Header* (scanner metadata). An *Air* is a parameter file for spatial adjustment, and an *AirVector* is a set of such parameter files.

SwiftScript relies on XDTM as its data model and type system. XDTM is motivated by the fact that logically clean data structures are often represented by "messy" physical representations in odd formats and storage organizations. For instance, an fMRI study is physically represented in a nested directory structure, with metadata coded in directory and file names, and a *volume* is represented by two files located in the same directory, distinguished only by file name suffix [15]. While we can incorporate the complex file system layouts and file formats into application programs and workflow specifications, the resulting code is hard to read and maintain, and cannot easily be adapted to changes in representations. Such messy physical representations make program development, composition, and execution unnecessarily difficult.

We address this problem by using XDTM, which allows logical datasets to be defined in a manner that is independent of the datasets' concrete physical representations. XDTM employs a two-level description of datasets, defining separately via a type system based on XML Schema the abstract structure of datasets, and the mapping of that abstract data structure to physical representations.

A dataset's *logical structure* is specified via a subset of XML Schema, which defines primitive scalar data types such as Boolean, Integer, String, Float, and Date, and also allows for the definition of complex types via the composition of simple and complex types. We use a C-style syntax (as shown in the example) to represent type schema, which gets translated into XML Schema transparently, so users don't have to deal with XML Schema directly. Such logical structures are mapped to physical dataset access during execution time by a set of mappers.

```
1   type Image {}
2   type Header {}
3   type Volume { Image img; Header hdr; }
4   type Run { Volume v[ ]; }
5   type Air {}
6   type AirVector { Air a[ ]; }

7   (Volume ov) reorient ( Volume iv, string direction, string overwrite)
8   {
9       app {
10              reorient @filename(iv.hdr) @filename(ov.hdr) direction overwrite;
11      }
12  }

13  (Run or) reorientRun (Run ir, string direction, string overwrite)
14  {
15      foreach Volume iv, i in ir.v {
16              or.v[i] = reorient(iv, direction, overwrite);
17      }
18  }

19  (Run resliced) fmri_wf ( Run r) {
20      Run yroRun = reorientRun( r , "y", "n" );
21      Run roRun = reorientRun( yroRun , "x", "n" );
22      Volume std = roRun.v[1];
23      AirVector roAirVec = alignlinearRun(std, roRun, 12, 1000, 1000, "81 3 3");
24      resliced = resliceRun( roRun, roAirVec, "-o", "-k");
25  }

26  Run bold1<run_mapper;location="fmridc/functional_data/",prefix="bold1">;
27  Run sbold1<run_mapper;location="fmridc/functional_data/",prefix="sbold1">;
28  sbold1 = fmri_wf (bold1);
```

**Figure 1 Sample fMRI Workflow**

## 3.3. Procedures

Datasets are operated on by *procedures*, which take typed data described by XDTM as input, perform computations on those data, and produce data described by XDTM as output.

Lines 7-12 in the example define what we call an *atomic procedure*, which specifies the interface to calling an executable program. The procedure *reorient* has typed input parameters *iv*, *direction* and *overwrite* (to the right of the procedure name) and one output parameter *ov* (to the left of the procedure name). The body of this particular procedure specifies that it invokes a program (conveniently, also called reorient) that will be dynamically mapped to a binary executable; this executable will be invoked at an execution site chosen by the Swift system at run time. The body also specifies how input parameters map to command line arguments. The notation *@filename* is a built-in mapping function that maps a logical data structure to a physical file name. In this case, it extracts the file name of input header and output header, which are then put in the command line to invoke the *reorient* program.

Lines 13-18 define a compound procedure *reorientRun*. A compound procedure calls one or more other SwiftScript procedures. The procedure takes in a run *ir* and applies the procedure *reorient* (which rotates a brain image along a certain axis) to each volume in the run to produces a reoriented run *or*. Because the multiple calls to *reorient* operate on independent data elements, they can proceed in parallel.

Lines 19-25 define another compound procedure *fmri_wf*, which itself calls a series of procedures including the *reorientRun* defined above. The definitions for the other procedures *alignlinearRun* and *resliceRun* are similar to *reoreintRun* and omitted here. This procedure essentially defines a simple four-step pipeline computation, using variables and/or datasets to establish data dependencies. It applies *reorientRun* to a run first in the *x* axis and then in the *y* axis, and then aligns each image in the resulting run with the first image. The program *alignlinear* determines how to spatially adjust an image to match a reference image, and produces an *air* parameter file. The actual alignment is done by the program *reslice*. Note that variable *yR*, being the output of the first step and the input of the second step, defines the data dependencies between the two steps.

As demonstrated in the example, procedures can be further called by other procedures, and thus constructing a sub-workflow within more complex workflows. The procedure abstraction allows arbitrary complex workflows to be easily composed from simple ones, and enables the sharing and collaborative development of complicated analysis procedures.

## 3.4. Iterations

The *foreach* statement at lines 15-17 defines iteration over the dataset *ir*. Iterations and procedural abstractions can be used to exploit repetitive patterns in workflows (and programs in general). While procedures are used to apply repetition in a lexical sense, based on the semantics of the problem (it is known at workflow design time where these repetitions occur), iterations apply to repetitions based on dynamic data (repeatedly executing a block of code for each datum in a collection).

Using iterations and procedures has impact on both workflow specification and execution. Exploiting repeated patterns and redundancy in workflows can lead, depending on the problem of course, to significant reductions in both code size and memory usage for the workflow system. As stated by Emmerich et al., the BPEL specification for a 4x200 flow leads to a size of 5MB [10]. In Swift, however, a few lines with nested *foreach* statements (for example, a nested set of iterations that applies the program *reorient* to each volume in a whole *Study*) can express hundreds of thousands of parallel tasks.

## 3.5. Dataset Mapping

The logical datasets defined in SwiftScript need to be mapped to concrete physical storage when the script is executed. Line 26 and 27 specify the mappings for the input dataset *bold1* and the output dataset *sbold1* respectively. Each mapping specification consists of three pieces of information: the type of the dataset (*Run* in this case), the mapping descriptor (*run_mapper*), and mapping parameters (*location* and *prefix*). What *run_mapper* does is that it goes into the specific location, finds the pairs of image and header files that share the specific prefix and have suffixes ".hdr" and ".img" respectively, and returns a *run* containing an array of *volumes*.

We define a standard mapping interface, and data providers implement the interface to support access to various data representations. A mapping descriptor provides the pointer to a mapping implementation. We also provide default mapping implementations for string mapping, file system mapping, and CSV (comma separated-value) files.

## 3.6. Dynamic Workflow Structure

The mapping process is activated when a SwiftScript is executed by the Swift system, and the datasets need to be evaluated. A *foreach* statement must be expanded prior to execution into a set of nodes: one per component of the compound data object specified in the *foreach*. This expansion is performed at runtime: when a *foreach* node is scheduled for execution, the appropriate mapping function is called on the specified dataset to determine its members (a set of *volumes* in this case), and for each member of the dataset identified (for each *volume* in the *run*) a new job (*reorient*) is created.

This dynamic expansion mechanism also allows us to switch input datasets (e.g., from a run of 10 volumes for testing purpose, to another run of 500 volumes for production) without changing either the workflow definition or the execution system, since the mapper discovers the dataset components at run time.

Dataset mapping not only abstracts away the complexity of dealing with "messy" physical representations, but also enables us to represent workflows whose structures can only be determined based on run-time data, a capability that is absent from most workflow systems based on static workflow structures. We illustrate this unique feature using a photo-realistic montage application [17][21].

Montage is a suite of software tools developed to generate large astronomical image mosaics by composing multiple small images. The typical process involves the following key steps:

- **Image projection**
  - Re-project each image into a common coordinate space
- **Background rectification**
  - Calculate a list of overlapping images
  - Perform image difference between each pair of overlapping images
  - Fit difference images into a plane
  - Background correction
- **Image Co-addition**
  - Optionally divide a region into a grid of sub-regions, and co-add the images in each region into a mosaic
  - Co-add the processed images (or mosaics in sub-regions) into a final mosaic

In the background rectification stage, a list of overlapping images first needs to be calculated, and then the image difference program is applied for each pair of the overlapping images. The specification of such dynamic workflows has turned out to be impossible for workflow systems based on static workflow structures, including our precursor – the virtual data system, and Condor DAGMan. It is also absent from other workflow systems (such as BPEL 2.0) that have iteration support but lack dataset mapping capability, because of the list of overlapping images can not be determined until the calculation is performed. Deelman et al. [17] approached the problem by developing specific C code (around 1200 lines of code) to call out the montage programs and generate the static workflow structure for each input image set.

Our XDTM mechanism affords us to represent the workflow in a compact and elegant manner. The list of the overlapping images contains the following columns:

*cntr1 cntr2 plus minus diff*

Where cntr1 and cntr2 are the image numbers, and plus, minus the image file names, and diff the difference image file name. In Figure 2 we show an example of the columns and rows in such a file. The SwiftScript program that maps the file content into a logical dataset and then iterates over the dataset is shown in **Figure 3**.

| cntr1 | cntr2 | plus | minus | diff |
|---|---|---|---|---|
| int | int | char | char | char |
| 0 | 91 | p_990427s-j1190056.fits | p_990427s-j1190044.fits | diff.000000.000091.fits |
| 0 | 191 | p_990427s-j1190056.fits | p_990427s-j1200221.fits | diff.000000.000191.fits |
| 0 | 441 | p_990427s-j1190056.fits | p_990427s-j1180221.fits | diff.000000.000441.fits |
| 1 | 95 | p_000704n-j0410209.fits | p_000704n-j0340056.fits | diff.000001.000095.fits |
| 1 | 246 | p_000704n-j0410209.fits | p_000704n-j0420056.fits | diff.000001.000246.fits |
| 1 | 1036 | p_000704n-j0410209.fits | p_000704n-j0410220.fits | diff.000001.001036.fits |
| 2 | 3 | p_980415s-j0620021.fits | p_980415s-j0610257.fits | diff.000002.000003.fits |
| 2 | 211 | p_980415s-j0620021.fits | p_980415s-j0610245.fits | diff.000002.000211.fits |
| 2 | 493 | p_980415s-j0620021.fits | p_980415s-j0620033.fits | diff.000002.000493.fits |
| 2 | 739 | p_980415s-j0620021.fits | p_980415s-j0630257.fits | diff.000002.000739.fits |
| 2 | 772 | p_980415s-j0620021.fits | p_980415s-j0630245.fits | diff.000002.000772.fits |

**Figure 2 The File Content of a Montage Overlapping Image List**

The program defines a logical structure *DiffStruct* corresponding to the columns, and uses a standard *csv_mapper* to map from the calculated overlapping list *diffsTbl* into a list of such logical structures, and then we can iterate over the list and apply difference program to each of the image pairs. The parameter *header* to the *csv_mapper* indicates that the file contains metadata (the column names), and *hdelim* indicates that they are delimited by character "|". The resulting SwiftScript program for the whole montage process is just 92 lines and it includes every stage and can adapt to any input image set without any code change.

SwiftScript also supports other dynamic features such as conditional execution, which are also used in the montage workflow specification to determine the number of sub-regions to co-add based on input data size. The details are omitted here due to space limitation.

```
type DiffStruct {
        int cntr1;
        int cntr2;
        Image plus;
        Image minus;
        Image diff;
}

// table of overlapping images
Table diffsTbl = mOverlaps ( projImgTbl );

DiffStruct diffs[]<csv_mapper; file=diffsTbl, skip=1, header=true, hdelim=" |">;
foreach d in diffs {
            Image image1 = d.plus;
            Image image2 = d.minus;
            // call the difference program with the two overlapping images
            Image diffImg = mDiffFit(image1, image2);
}
```

**Figure 3 Montage Workflow Specification (excerpts)**

## 3.7. Code Size Reduction

The primary focus of SwiftScript's design is to provide a compact and easy to code workflow notation. As an approximate measure of the claim, we compare in Table 1 the lines of code needed to express five different fMRI workflows, coded in SwiftScript, with two other approaches, one based on ad-hoc shell scripts ("Script," able to execute only on a single computer) and a second ("Generator") that uses PERL scripts to generate explicit DAG style pre-XDTM SwiftScript. With the script approach, the programmer must take care of locating the specific image files in directories, putting each of the file names into a program invocation, and tracking the outputs of each invocation. The generator approach works in similar ways in that it needs to locate the files; defines the invocations on each individual file, and also constructs the data dependencies between different analysis stages explicitly. The new SwiftScript programs are smaller and more readable—and also provide for type checking, provenance tracking, parallelism, and distributed execution that come with the virtual data system.

**Table 1 Lines of Code with Different Workflow Encodings**

| Workflow | Script | Generator | SwiftScript |
|---|---|---|---|
| GENATLAS1 | 49 | 72 | 6 |
| GENATLAS2 | 97 | 135 | 10 |
| FILM1 | 63 | 134 | 17 |
| FEAT | 84 | 191 | 13 |
| AIRSN | 215 | ~400 | 37 |

We also compare the code size of our SwiftScript program with the MPI approach in the Montage example. The reason that we can conduct such comparison is that the Montage application provides conventional sequential programs for processing individual images. For instance, the program *mProjectPP* performs fast projection on an image. The application also provides MPI programs that invoke sequential programs on multiple processors. An example is the *mProjExecMPI* program, which reads a list of image files and calls the *mProjectPP* program simultaneously on multiple processors, each processing one of the images. The MPI program is coded in C++ and the total number of lines is 950. In the SwiftScript program, we achieve the same parallelization by first declaring the function interface for *mProjectPP*, and then defining a compound procedure *mProjectBatch* that iterates over a list of images, and applies *mProjectPP* to each of the images. The total number of lines involved is 15. In addition to significant code size reduction, our high level SwiftScript specification also does not require any specific parallel directives such as the initialization of multiple processors, the assignment of processor ranks and the tracking of success and failed jobs etc. The SwiftScript program also has the advantage that it can be scheduled to execute on various number of computing resources across multiple clusters. We will show in the evaluation section that the Swift system also achieves comparable performance to MPI execution in running the Montage workflow.

## 3.8. The Swift Runtime System

The Swift runtime system (see **Figure 4**) is a scalable environment for efficient specification, scheduling, monitoring and tracking of SwiftScript programs. The system consists a SwiftScript compiler that compiles SwiftScript programs into abstract computation plans, an execute engine built on CoG Karajan and a set of libraries and tools for the scheduling and execution of the computation plans, a provenance tracking tool that records the execution process, and resource provisioning mechanism for submission to diverse computation and storage resources. We are still in the process to integrate the virtual data catalog (VDC) [42] as first introduced and implemented in the virtual data system to store the program, provenance, and metadata information.

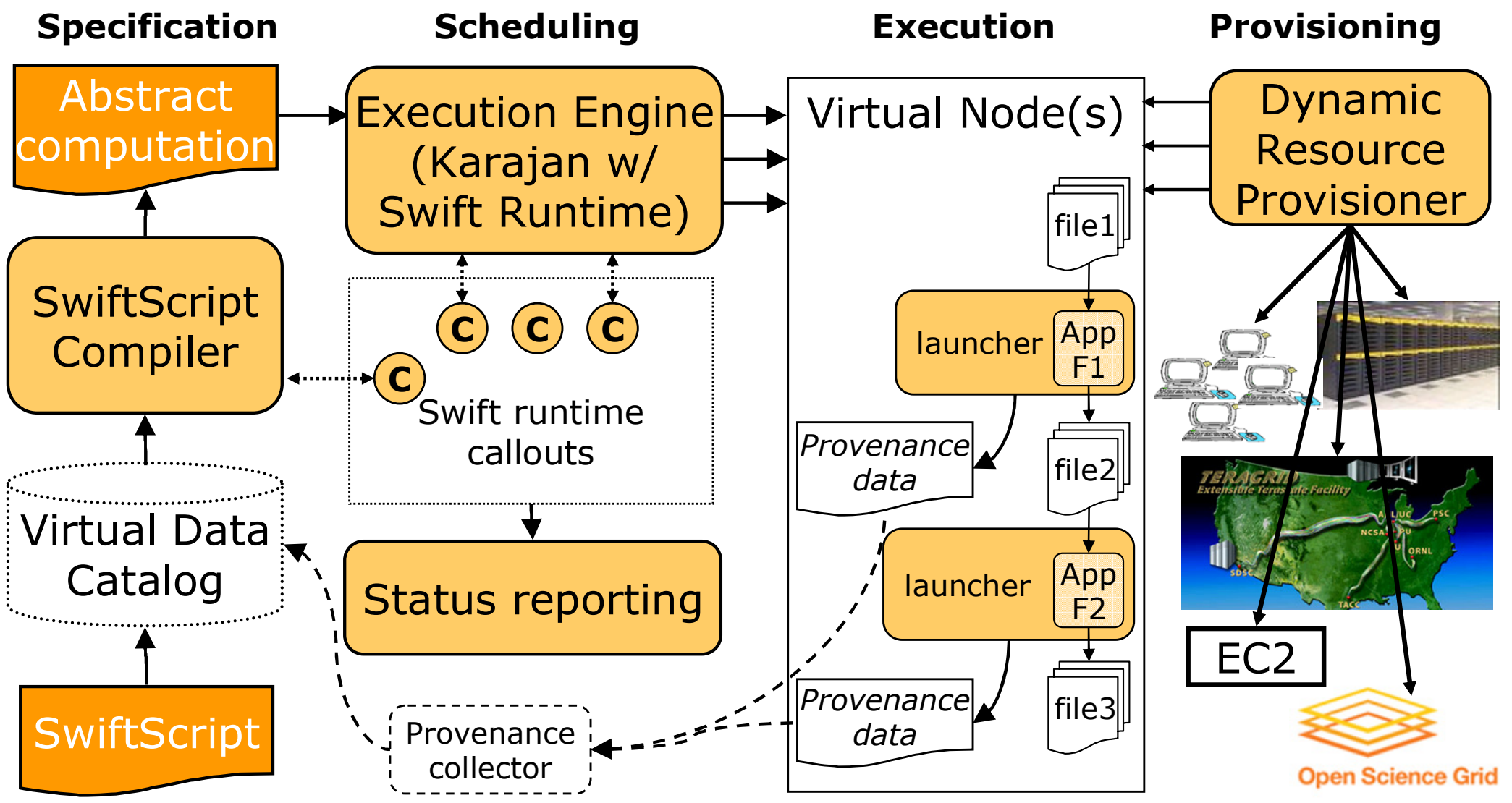

**Figure 4 The Swift Runtime System**

## 3.9. Program Compilation

SwiftScript programs are compiled into abstract computation plans that can be interpreted by the execution engine. SwiftScript borrows a number of concepts used in generic programming languages, some of which require support in the underlying execution language (such as conditional execution), and some of which benefit from support in the execution language (iterators and procedures). We choose CoG Karajan as the underlying execution engine, and Karajan scripts as the representations for execution plans. Karajan provides a set of abstractions (and corresponding libraries and implementations) for file operation, data transfer, job submission, and Grid services access. Such operations can be organized using Karajan language constructs such as sequential and parallel execution, sequential and parallel iterations, conditional execution and functional abstraction etc. We extend Karajan with specific libraries to support the XDTM type system and logical dataset manipulation, adapters to access legacy VDS components (for instance, the site catalog that keeps information about each execution site), mappers for accessing heterogeneous physical data storage, and also with fault tolerance mechanisms.

In SwiftScript programs are driven by data dependencies. Since explicit dependency analysis and translation into sequential and parallel operations would be very difficult, and the power and flexibility provided by path specifications and expressions in SwiftScript further complicate the process. We choose therefore to utilize existing future and event mechanisms in Karajan which allow us to bypass any dependency analysis and effectively let the engine dynamically figure out the dependencies. Karajan is not a data-flow oriented language, however it contains constructs that can facilitate data-flow oriented programming. Conceptually data-flow can be seen as a system where parts of the computation progresses based on the availability of data that flows between them. The data flow can be implemented in strict multi-threaded languages using *Futures* [13][20] (similar to single-assignment variables, which are constructs used for synchronization in concurrent programming languages [7]), a feature that Karajan supports. A *Future* is a mechanism to provide a placeholder for values that will be *resolved* asynchronously. The *Future* provides a way to ask for the actual value. *Two situations are possible*:

1. *The Future has resolved already. That is, the method calculating the value has finished, and stored the result in the Future. Asking for the Future's value will immediately give the result.*
2. *The Future has not resolved. In this case, the current thread is blocked until the Future resolves. The current thread is effectively synchronized with the thread calculating the value.*

So essentially we treat all computations as parallel and the future mechanism establishes the dependencies between them thus constructing the workflow structure dynamically at run time.

## 3.10. Scheduling

Karajan uses lightweight threading techniques to instantiate and dispatch tasks (It represents active tasks as “lightweight threads”, i.e. not a conventional Java thread but some brief description of an executable task) and thus can execute large-scale task graphs. One important source of memory utilization in general is the statically allocated stack space needed by threads and processes. Due to the architecture of current operating systems, threads and processes must statically allocate sufficient memory to handle the most demanding tasks, yet promote wasteful usage in the case of less demanding tasks. Threads also consume other operating resources which limit the number of total concurrent threads that can be started. But threads and/or processes are not entirely necessary for the execution of workflows. They serve the purpose of time-sharing processors. However, in the case of, for example, Grid job submissions, most of the time is spent waiting for notifications while a job is executing remotely. This wait period does not use the processor, but instead consumes memory (and other OS resources) for the very purpose of preserving state inherent to the threading mechanism itself. In other words, the necessity for time-sharing a processor exists only in short phases of Grid tasks. In addition, those relatively short phases, being processor intensive, do not benefit from time-sharing the processor. Consequently the use of threads is largely unnecessary for the coordination of Grid tasks.

Karajan addresses the problem of thread memory consumption by employing lightweight threading and asynchronous libraries (where possible). Lightweight threads allow dynamic allocation of only the strictly necessary memory for preserving task/workflow state information. The advantage of using lightweight threading (event-based systems) and asynchronous I/O has been previously studied in CML [31], SEDA [37], and in the Erlang language [3], originally developed by Ericsson to support scalable, distributed, and fault-tolerant applications (in particular telecommunication systems).

## 3.11. Execution

Intermediate abstract execution plans represented in customized Karajan scripts are interpreted and dispatched by Karajan onto execution sites. We call these plans abstract because executions can be carried out in a location independent fashion: tasks are dispatched to virtual nodes that can be bound to different kinds of computing facilities varying from personal desktops and clusters to multi-site Grid environments. Swift uses *just-in-time* planning to determine at runtime the actual execution processes such as site selection, data stage-ins and stage-outs, and runtime error checking. A set of callbacks allow customized functions to determine where to dispatch tasks, how to group tasks to increase granularity, and/or when and how to perform data staging operations etc. The derivation process is also captured by the system and stored in the virtual data catalog.

Swift can schedule the execution of a program on different compute resources based on an abstract provider interface in CoG Karajan. The interface defines the processes involved in a job execution such as *submit*, *suspend*, *resume*, and *cancel*. If we implement the interface for a computation resource, for instance, the local host, then we can submit jobs to the local host provider.

Karajan provides the implementations for the following execution resources:

- Local host, jobs can be submitted to the local host, for instance, a workstation, a laptop, etc.
- Cluster scheduler submission (PBS or Condor), jobs can be submitted to a computer cluster managed by PBS or Condor.
- GRAM job submission [11], jobs can be submitted to Globus GRAM gatekeeper, or the GRAM-WS service released in Globus Toolkit 4.

- Falkon [30] lightweight task execution service that can dispatch and execution fine grained tasks using mulit-level scheduling and a streamlined task dispatcher.

There is a configuration file that specifies which of these providers to submit to. Thus the same SwiftScript program can be configured to execute either on a local workstation, a LAN cluster, or on multi-site Grid environments.

## 3.12. Fault Tolerance

Swift addresses reliability issues at several levels. At the software development level, its type checking capabilities allow it to identify potential problems in a program prior to execution. Its support for virtual nodes makes it easy to first test a program using a local host submission mechanism with a small set of datasets, and then move to larger problems and execution sites.

Swift provides exception handling and retry mechanisms to deal with failures that may occur at different levels. Transitory problems are recovered by retrying the faulty tasks (for instance, retrying a transfer if a GridFTP server is busy). Host-level faults (where a resource exhibits problems with unknown duration) are dealt with by rescheduling a task on a different site. Host-level faults can also be dealt with downstream, for example in Falkon. Common problems can be identified and tagged accordingly in order for Falkon to avoid particular faulty hosts when known problems arise. For example, a common problem that happens when running large scale applications on many CPUs with data on shared file systems is the “Stale NFS handle” error, which simply means that the shared file system mount is in an inconsistent state (likely due to the shared file system’s I/O servers being overloaded); this error implies that all tasks that execute on the particular compute node that need data from the shared file system will fail. With Falkon’s ability to suspend faulty hosts (for a specified amount of time), tasks can be rescheduled on other hosts in the same site, and only after a certain number of repetitive failures in the same site, does Falkon pass on the error back to Swift in order to have the task rescheduled at a different site.

Swift also keeps a restart log, allowing it to resume the state of a computation in case of premature termination (e.g., caused by a machine reboot). The restart log is similar to a Condor rescue DAG [6], except that Condor tags jobs that are finished, whereas we log datasets that are successfully produced. The reason is that the Condor DAG is control- flow based, which keeps track of job status; where Swift evaluates workflows by data availability so it makes more sense to track data in our case. This logging mechanism can resume a workflow from where it was stopped, since the workflow evaluation mechanism is based on data availability, and when we restart the execution of a workflow, we mark the logged datasets as being computed and available, then the stages dependent on such datasets can continue executing, thus correctly resuming the execution of the workflow. We have tested restartability by repeatedly interrupting program execution, and verified that our programs continue from where they were interrupted. We also note some (good) side effects to this mechanism: (a) new inputs can be added after a computation has been run for some time, and once we restart the computation, the system is able to figure out that these new inputs are present and not processed, and thus schedule their executions. (b) We can make modifications to a program and restart it, as long as the modifications do not affect prior data flows. This effect is useful for debugging and testing purposes.

## 3.13. Optimization

Swift also supports advanced Grid scheduling and optimization, such as pipelining, clustering, and load balance etc.

**Pipelining** refers to the ability to optimize execution by executing dependent iterations incrementally. Swift is built on data-driven mechanism, so once an item in a collection is processed, any processes that are dependent on that data item can proceed right away without waiting for the whole collection to finish. This advantage comes for free with the future mechanism, as a particular piece of data can be used as soon as it is available (resolved), regardless of the lexical and structural intricacies of the program.

**Clustering/Partitioning** is a solution that tries to reduce costs incurred by various phases in the task submission process. It is based on the observation that the initialization phase is computationally intensive and for large numbers of short jobs it can introduce a high overhead. It attempts to identify simple structures in a workflow and combine them into a single task. Partitioning can also be useful in scheduling closely related tasks on the same resource in order to minimize data movement between resources. Since a SwiftScript program is dynamically evaluated, we do not need the knowledge of the whole workflow graph for partitioning, we instead introduce small delays (a clustering window) in task submission with the purpose of accumulating independent tasks. This solution would identify groups of possibly independent jobs in the graph based on the time the system attempts to submit them, and group them without exceeding a pre-set bundle size (or total execution time). Compared with static partitioning engaged in Pegasus, our dynamic clustering does not need prior knowledge of the workflow graph, and it can better adapt to changing conditions such as job dispatching rate and available compute resources, but in the mean time its lack of global knowledge also makes it harder to perform higher level partitioning, such as vertical clustering (group jobs in multiple consecutive stages). The Falkon lightweight execution service provides better granularity control than clustering as the dispatch overhead per task is small enough that we can process each task individually. However, in a cluster environment where Falkon is not installed, clustering can be used to improve the overall throughput over the more heavyweight schedulers (i.e. PBS, Condor, etc).

**Load Balancing**: Swift programs can be scheduled and executed on multiple Grid sites. Swift firstly chooses valid sites for the execution of a specific task by checking site availability, and also making sure the required applications for the task are installed on those sites. It then applies some heuristics based on site responsiveness to dispatch jobs. Basically, each site is given a score associated with how fast and reliable it turns jobs around, the score is increased when jobs run successfully and decreased upon exceptions. Jobs are dispatched to each site proportional to its score. More sophisticated proactive strategies can be adopted using Swift call-outs to schedule jobs based on, for instance, site resource monitoring information.

## 3.14. Provenance Tracking

When a job is run on a computing resource (e.g., a remote site or local host), we initiate its POSIX process execution using an application-launching tool called *Kickstart* [36]. Kickstart serves two purposes—it provides a uniform way for Swift to pass arguments to an execution site, and it captures detailed information about the actual execution environment and the application's behavior. This information is returned after execution to the submitting host as an *invocation document* and is stored in the VDC. The invocation document contains environmental details (such as host name, IP address, current working directory, and the complete set of environment variables), information about the behavior of the application (including its exit code and signals), and the application's resource usage (including its system time and user time, paging and swapping activity, and OS context switches). Virtually all information available about the process through standard POSIX/Linux end-user interfaces is captured. Such information has proven invaluable in debugging and monitoring distributed applications, and in recording and querying the provenance of the data derived by Swift applications [28][42].

Now that we have SwiftScript to specify large complex parallel computations, and we have the Karajan execution engine to efficiently and reliably dispatch the large number of jobs, we need to make jobs execute efficiently on distributed Grid resources. As we have already mentioned, jobs can be dispatched to a local provider where jobs run on a local workstation or laptop. However, this is only suitable for small test cases to verify that the workflows run as intended. To run larger and production level experiments, Swift dispatches (usually via GRAM) jobs to parallel computing clusters where hundreds and thousands of computers can be used. In the latter case, we rely on batch schedulers running on each execution site to handle the jobs dispatched to them. However, batch schedulers (such as Condor and PBS) commonly used to manage parallel computing clusters are not typically configured to enable easy configuration of application-specific scheduling policies. In addition, their sophisticated scheduling algorithms can be relatively expensive to execute, and do not scale well to long lists of queued jobs. Applications that require the rapid execution of many small tasks often do not perform well. For instance, the *reorient* process in the fMRI workflow can involve hundreds or thousands of volumes, but each reorientation only takes no more than a few seconds, and the Montage image difference process shows the same pattern. Such processes usually incur large overhead and low throughput using a conventional batch scheduler. In order to address this issue, we have developed and integrated Falkon, a lightweight task execution service that can dispatch and execution fine grained jobs an order of magnitude faster than conventional batch schedulers.

# 4. THE FALKON EXECUTION SERVICE

Falkon is a Web Services-based system whose main goal is to support the efficient execution of large numbers of small (fine-grained) tasks in batch scheduled environments that are typically found in most production Grids.

To enable the rapid execution of many tasks on compute clusters, we have developed Falkon (see **Figure 5**) [30], a Fast and Light-weight tasK executiON framework. Falkon integrates (1) multi-level scheduling to separate resource acquisition (via, e.g., requests to batch schedulers) from task dispatch, and (2) a streamlined dispatcher. Falkon avoids the need for reconfiguring existing queues (one possible approach that is not practical in real deployed systems) by submitting one or more requests for nodes and then deploying on those nodes software that implements a domain-specific scheduling strategy. By thus embedding a new scheduler inside the old, it separates the two tasks of resource *provisioning* (acquiring the resources needed for a computation) and *scheduling* (mapping tasks to those resources). Once the resources are acquired, they can be used repeatedly for executing jobs, without incurring the overhead for allocating resources for each job. Resources can also be allocated or de-allocated dynamically with regard to a workload. In Falkon, resource provisioning is performed by the DRP (Dynamic Resource Provisioning) component [29][30]. DRP can be useful when either resource demand or resource availability varies during the course of program or workload execution, in which case it can improve execution times and/or the efficiency of resource utilization. Falkon's integration of multi-level scheduling and streamlined dispatchers delivers performance not provided by any other system. We describe Falkon architecture and implementation, and present performance results for both microbenchmarks and applications.

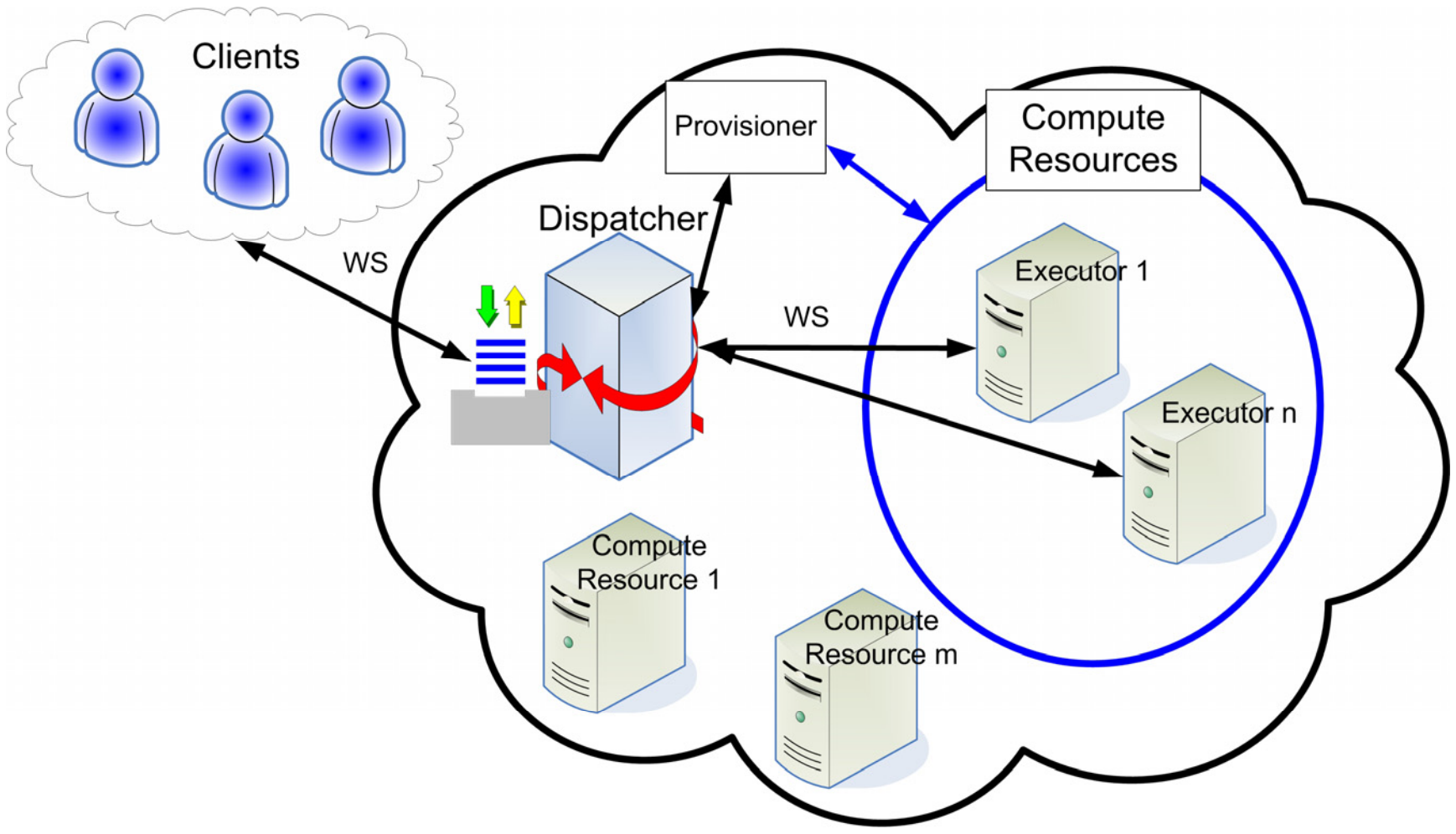

**Figure 5 Falkon Service Architecture**

Falkon is built on top of a Web Services-based architecture that offers applications a simple and efficient interface into a complex set of Grid resources. Falkon provides a Web Services interface for users to submit job execution requests. The jobs are first queued at the service. Falkon then calls out to DRP, which uses WS-GRAM to allocate resources in any Grid site after which the pool of resources can be maintained, increased, and decreased based on the load (queue length) of the Falkon service and the policies defined within DRP. For instance, DRP will try to allocate more workers if jobs start piling up in the service queue, and a worker can be configured to automatically de-register itself from the service if it has been idle (no jobs dispatched to it) for a certain time. The acquired resources register themselves with the Falkon service after which jobs in the service queue will be dispatched to the resources based on scheduling decisions. The Falkon service is unique in the sense that it can achieve relatively high throughput in the range of hundreds to thousands of tasks per second when running fine grain tasks, with a per task overhead measured in milliseconds. This unique feature of Falkon in combination with the dynamicity of provisioning can enable a wide range of applications to run significantly more efficiently, due to efficient task dispatch and the dynamic resource provisioning that simplify the compute resource management at the applications.

Micro-benchmarks show that Falkon can achieve throughput up to 487 tasks/second in a general case where each task is handled separately (with throughputs exceeding 2500 tasks/sec with techniques similar to the clustering defined earlier), manage up to 54,000 executors, and scale to 1.5 million queued tasks.

We conducted similar benchmarks on Condor and PBS to gain insight into how Falkon performs as compared with those systems. We set the number of resources to 32 nodes (64 processors) and measured the time to complete 64 jobs of various task lengths (ranging from 1 sec to 16384 sec). We then calculated the efficiency of resource usage as:

$$E = S_p / S_I$$

$S_p$ is the practical measured speedup, where $S_I$ is the ideal speedup, which equals to the number of processors being used. The maximum efficiency thus is 100%.

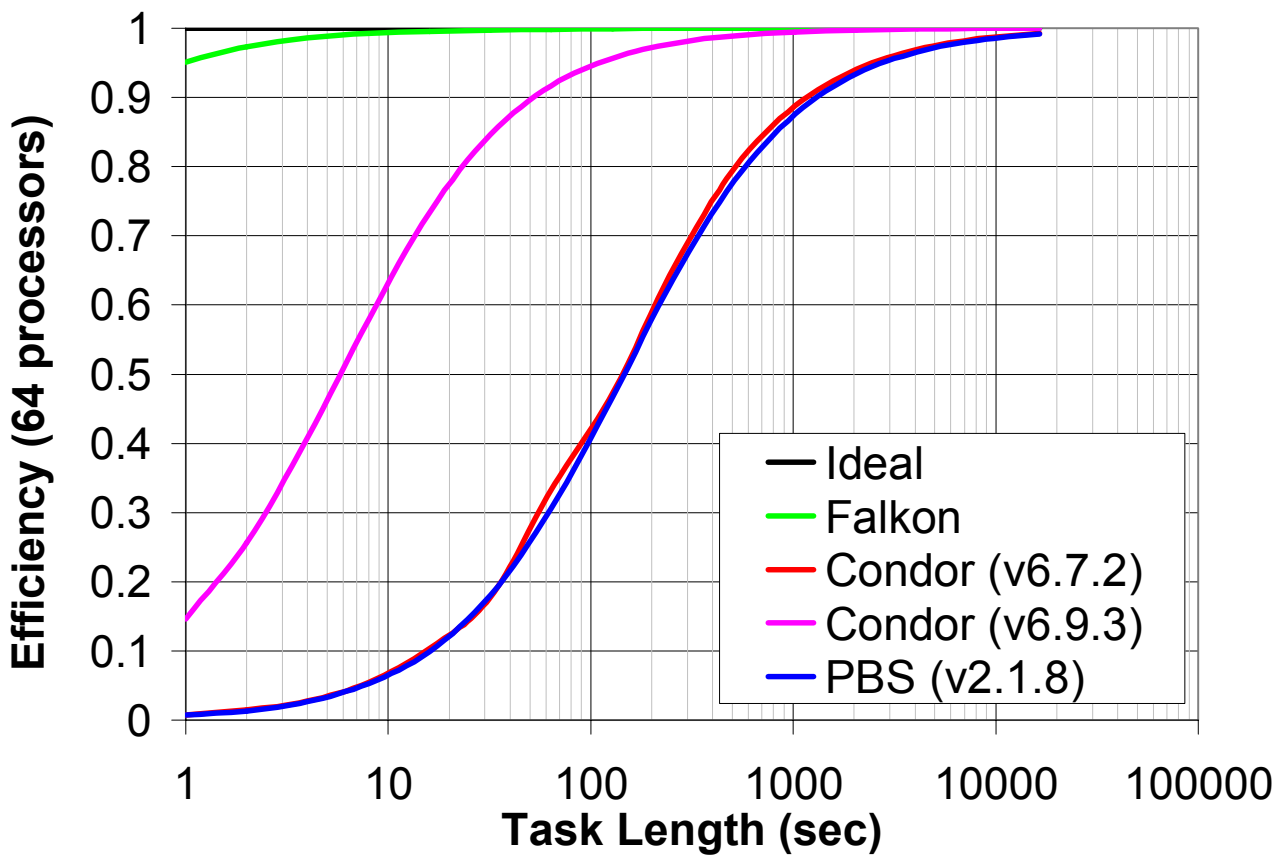

**Figure 6: Efficiency of resource usage for varying task lengths on 64 processors comparing Falkon, Condor and PBS**

shows Falkon's efficiency to be 95% with 1 sec tasks and 99% with 8 sec tasks. In contrast, both PBS (v2.1.8) and Condor (v6.7.2) have an efficiency of less than 1% for 1 sec tasks and require about 1,200 sec tasks to get 90% efficiency and 3,600 sec tasks to get 95% efficiency. They can only achieve 99% efficiency with 16,000 sec tasks.

As both the PBS and Condor versions being tested on the experiment clusters (ANL/UC Teragrid site) are not the latest versions, we also derived the efficiency curve for Condor version 6.9.3, the latest development Condor version, which is claimed to have a throughput of 11 jobs/sec [39] (up from our measured 0.5 jobs/sec and the 2 jobs/sec reported by others [32]). Efficiency is much improved, reaching 90%, 95%, and 99% for job lengths of 50, 100, and 1000 seconds respectively. The results in  for Condor v6.9.3 are derived, not measured. We derived based on the achieved throughput cited in [39] of 11 tasks/sec for sleep 0 tasks. Essentially, we computed the per task overhead of 0.0909 seconds, which we could then add to the ideal time of each respective task length to get an estimated task execution time. With this execution time, we could compute speedup, which we then used to compute efficiency. Our derivation of efficiency is simplistic, but it allowed us to plot the likely efficiency of the latest development Condor code against the older production Condor code, the PBS production code, and Falkon. It should be noted that  illustrates the efficiency of these systems for a relatively small set of resources (only 64 processors), and that the efficiency gap will likely only increase as the number of resources increases (see **Figure** 7).  Also,  shows only the efficiency for a small number of jobs (64), which makes slower throughput scenarios (PBS and Condor) hard to achieve high efficiency due to the large time to dispatch the tasks to the CPUs.  We therefore generalize the results from  with the results in **Figure** 7.

We emphasize the need for such high throughputs by measuring the theoretical (assuming 1M tasks) resource efficiency at three different scales (small grid sites – 100 processors, medium size grid site – 1K processors, large Grid – 10K processors) for various throughputs (1, 10, 100, 500, 1K, 10K, 100K, and 1M tasks/sec). It is worth noting that current production LRMs (i.e. throughput of 1 task/sec) require relatively long tasks in order to maintain high efficiency. For example, even in a small Grid site with 100 processors, tasks need to be 100 seconds in duration just to get 90% efficiency; the task duration is increased to 900 seconds for a modest 1K processors, 10K seconds (~2.8 hours) for 10K processors just to maintain 90% efficiency. With throughputs in the range of 500 tasks/sec (which is obtainable with Falkon), the same 90% efficiency can be reached with tasks of length 0.2 seconds, 1.9 seconds, and 20 seconds for the same three cases.  It should be evident that the higher the throughput of tasks/sec that can be dispatched and executed over a set of resources, the higher the resource efficiency for the same workloads and the faster the applications turn-around times will be.

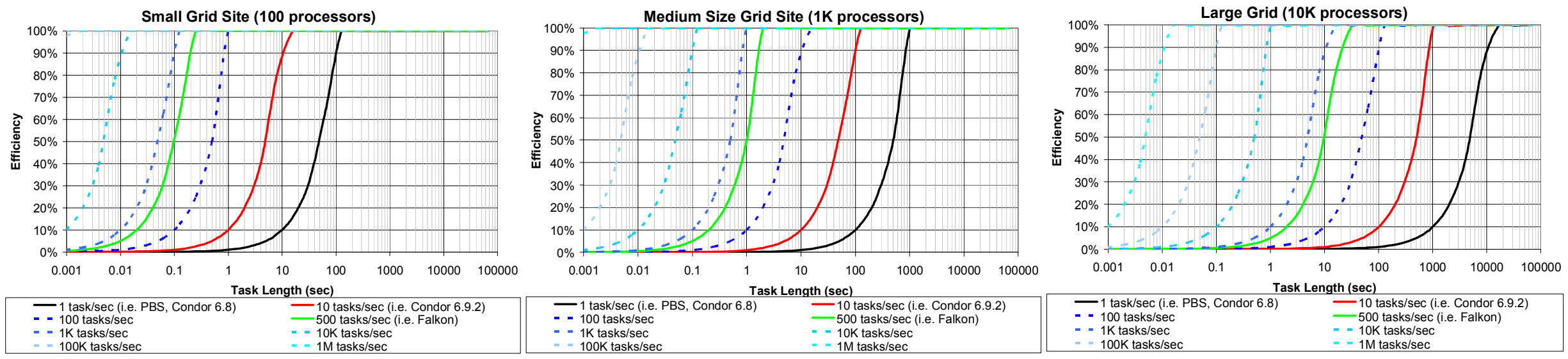

**Figure 7: Resource Efficiency for various resource scales and workload characteristics**

Finally, we wanted to investigate the effects of the task dispatch rates on the I/O throughput that could be achieved with Falkon and with Condor or PBS to and from the shared file system and compare that with the ideal I/O throughput without any provisioning, task dispatch, and scheduling overhead. The experiments involved the execution of tasks on 64 nodes which consisted of tasks with input and/or output with varying sizes from 1B to 1GB.  The shared file system in this case was GPFS and it was configured with 8 I/O servers to handle the traffic in and out of the shared file system. The results are shown in **Figure 8**. It is interesting to note that Falkon (primarily due to the high dispatch rates it can achieve) can achieve close to the ideal I/O throughput for both read performance.  On the other hand, Condor and PBS both require large (i.e. 1GB) input and/outputs per job in order to achieve the same level of performance that Falkon can achieve with significantly smaller input and outputs per job (i.e. 1MB).

For real applications, we show in the System Evaluation Section that Falkon can reduce application execution time significantly (up to 90%), as compared with other job submission mechanisms, such as GRAM plus PBS.

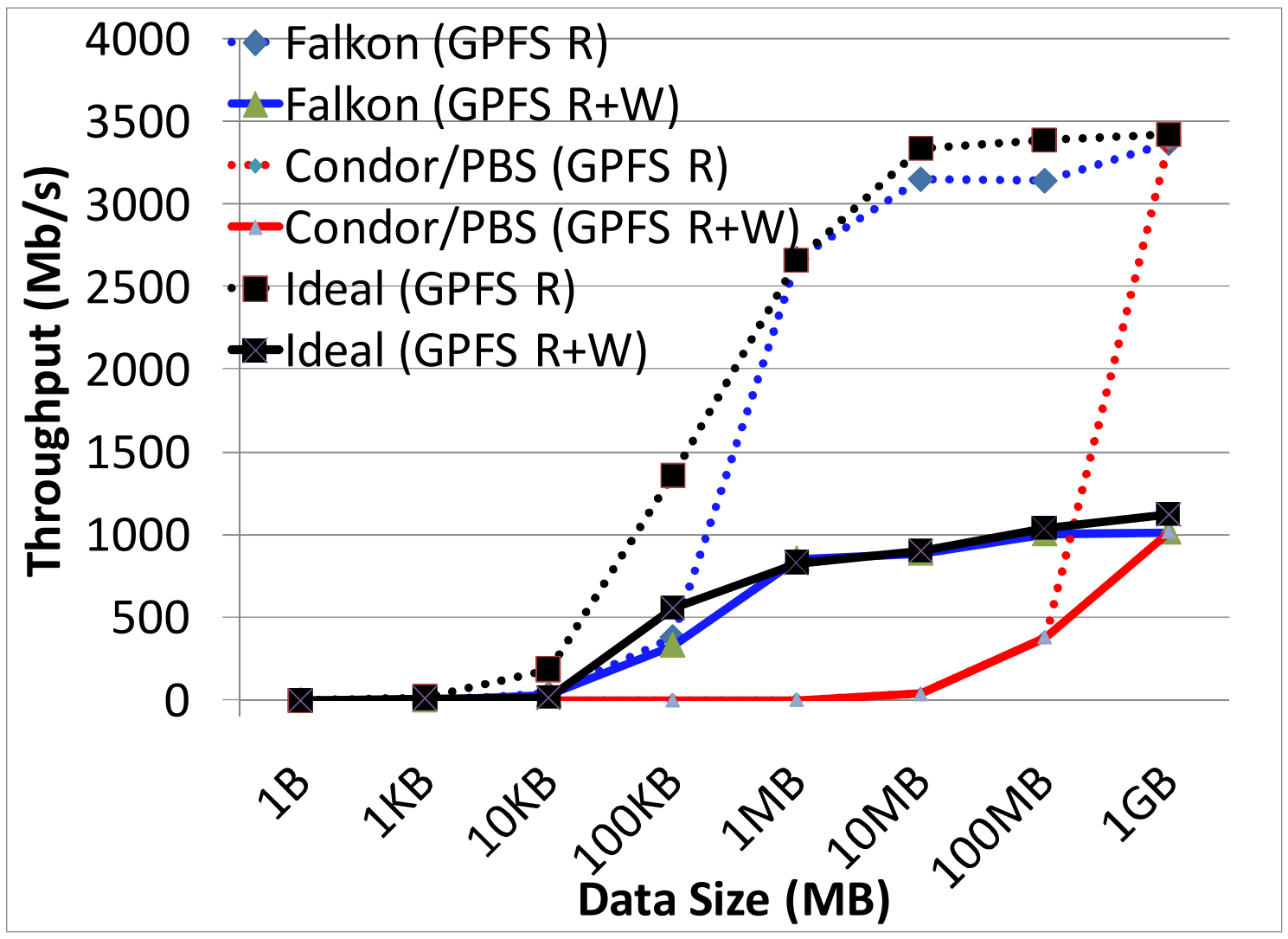

**Figure 8: The effect of task dispatch rates and how it affect I/O throughput**

# 5. SYSTEM EVALUATION

We evaluate various aspects of the Swift system using both synthetic workloads and real scientific applications, and we report the experiment results in the following sections. Unless otherwise specified, the experiments used a machine at the computation institute at the University of Chicago as the Swift submit host, and we used two computation clusters, one is the IA64 clusters at the ANL/UC TeraGrid site with 62 nodes, and the other the computation Institute Teraport cluster with 120 nodes. Both clusters had 1 Gb/s Ethernet connections and used PBS as the default local scheduler. We summarize the configurations of the test environment in **Table 2**. The ANL_TG cluster was used as the default execution site to which jobs were submitted.

**Table 2 Configurations of the Test Environment**

| Name | Type | Location | #nodes | Configuration |
|---|---|---|---|---|
| **UC_SUBMIT** | Submit Host | UChicago CI Workstation | 1 | Intel P4 2.4GHZ<br>1 GB Memory |
| **ANL_TG** | Execution Site | ANL/UC Teragrid Cluster | 62 | Intel Itanium 2<br>Dual Proc. 1.3 GHZ<br>4 GB Memory |
| | | | 98 | Intel Xeon<br>Dual Proc. 2.4 GHZ<br>4 GB Memory |
| **UC_TP** | Execution Site | UChicago CI Teraport Cluster | 120 | AMD Opteron<br>Dual Proc. 2.2 GHZ<br>4 GB Memory |

## 5.1. Scalability

We consider first the size of the parallel computations that can be executed correctly. Since Swift integrates Karajan, we measure the maximum number of tasks (or the maximum number of nodes in a task graph) that can be processed and dispatched with certain amount of memory in both systems. Figure 9 shows the scalability results. For the Karajan measurement, we used a simple script that creates multiple parallel threads; each of the threads merely waits for a certain amount of time (so all the threads can be created before any of them terminates). In doing so, we make sure that there is no extra overhead associated with each thread other than the information necessary for its creation.

For the Swift measurement, we used a synthesized workflow where we could set the number of nodes in the workflow and also the data dependencies (edges) between graph nodes. We set the Java virtual machine (in which the Swift scheduler runs) heap size to use certain amount of memory, and we kept adding nodes in the workflow until the system ran out of memory in processing the workflow to get the maximum number of nodes. The Karajan system is shown to be highly scalable, with 32 MB memory, it can create more than 40,000 threads. Each thread has a memory footprint of 800 bytes, which is the base line as each thread only contains the minimal information. Our Swift in general has much higher memory footprint (3.2 KB for each node, the size could vary with different workflow graphs and procedure definitions, as both data dependencies and input, output parameters need to be kept in memory), but can still support up to 4,000 threads with 32 MB memory, and 160,000 nodes with 1 GB memory. The reason each node takes more space is because we use futures for data synchronization, and each future object wraps a java dataset object. Also for each node, we need to store its associated procedure information, such as input, output names, types and values etc. Nevertheless, the lightweight Karajan thread leaves much space for further improvement of our Swift system.

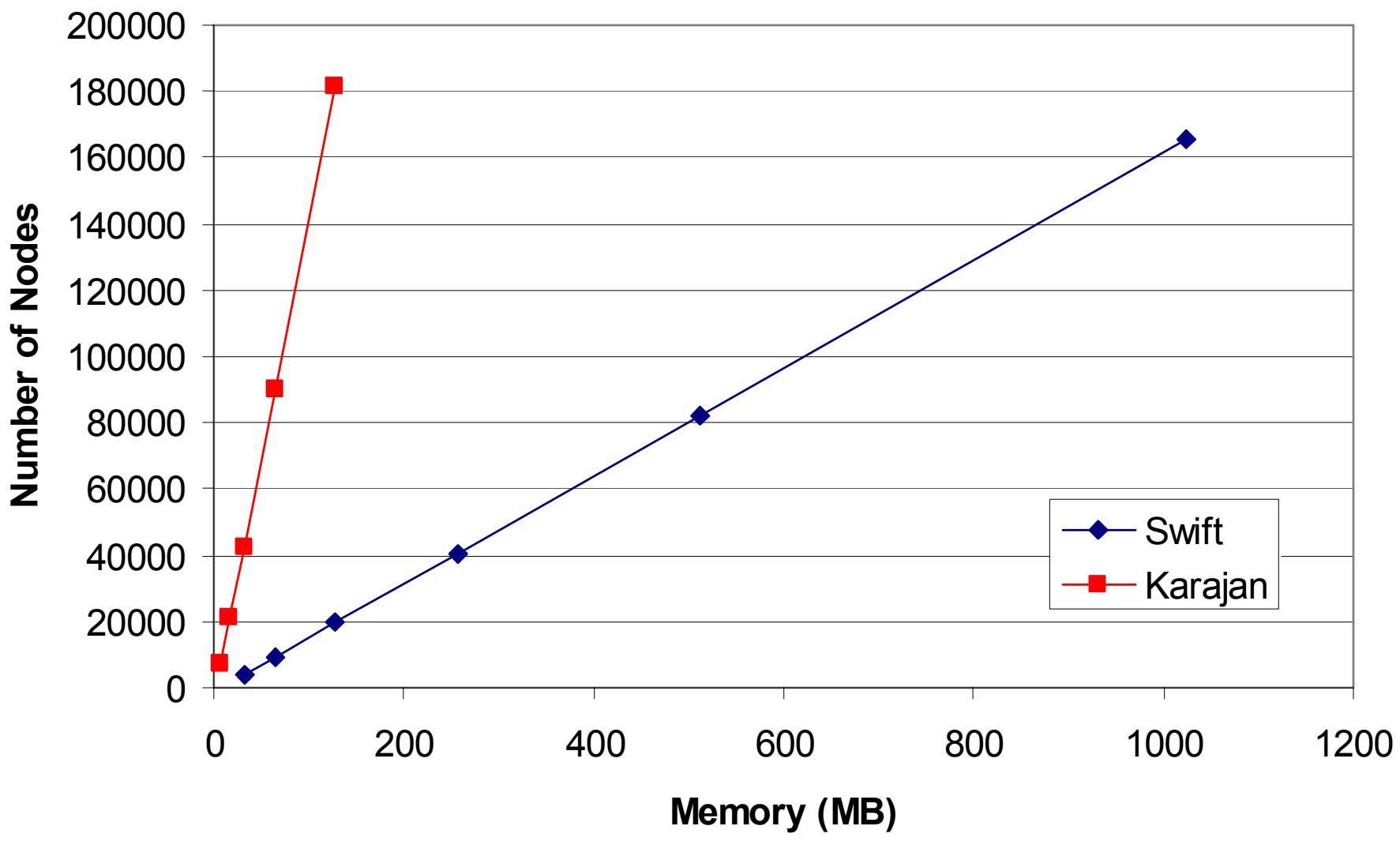

**Figure 9 Swift System Scalability**

## 5.2. Pipelining

Our SwiftScript language design allows us to implement pipelining across iterations (which is similar to the map operator in functional programming languages) for improved performance. In Swift, we realize pipelining using the future mechanism we adopt for evaluating data dependencies. Pipelining allows us to dispatch jobs more efficiently as they become ready.

We evaluate the effect of pipelining in our implementation and show the result In . On the top, we show the execution time for the fMRI workflow defined in **Figure 1** without pipelining, and on the bottom, the execution time with pipelining. The workflow operated on a 120-volume input dataset and involved four stages of 120 computations each. We can observe that the run without pipelining had to wait for all computations in one stage to finish before it could proceed to the next stage, thus the distinct start times for all four stages. While in the pipelined run, the computations between different stages were overlapped. Consequently, the execution time was reduced by 21% for the fMRI workflow with pipelining enabled.

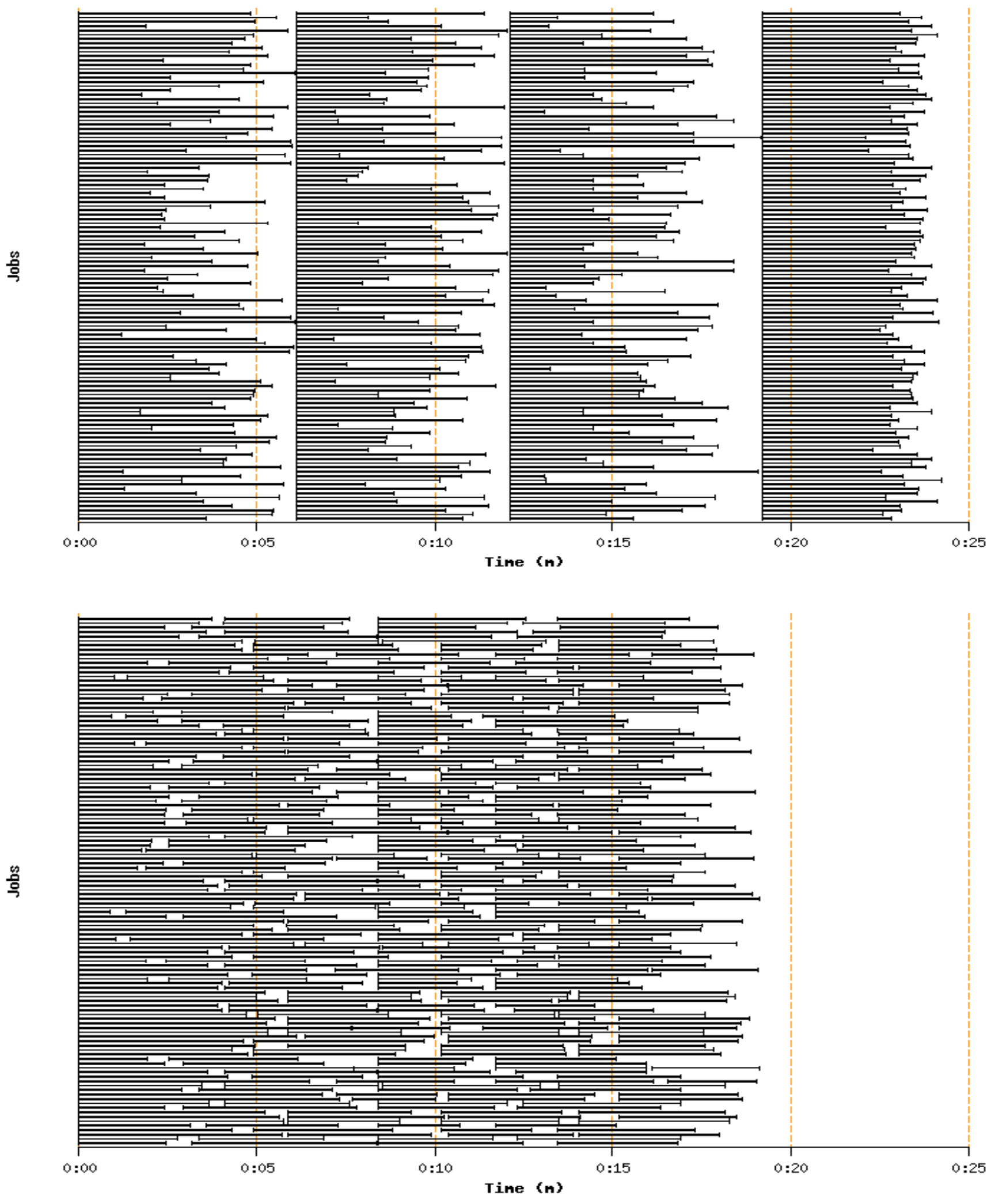

**Figure 10 The Pipelining Effect for the fMRI Workflow**

## 5.3. Load Balancing

Load balancing is achieved in Swift using heuristics based on site responsiveness (Each site has a score associated with its load and the turn-around rate for job submissions). We tested system load balancing also using the 120-volume input fMRI workflow, and we configured Swift to submit from UC_SUBMIT to both the ANL_TG and UC_TP clusters. Since UC_TP was in the same local area network as that of the submit host, and the processors on UC_TP were faster than those on ANL_TG, jobs dispatched to the UC_TP cluster completed faster, thus we can observe in Figure 11 that overall the system balanced the load across the two sites, ANL_TG got 218 jobs (in solid blue) out of the total number of 480, and UC_TP got the other 262 jobs (in solid gray), and relatively more jobs were scheduled on the UC_TP cluster.

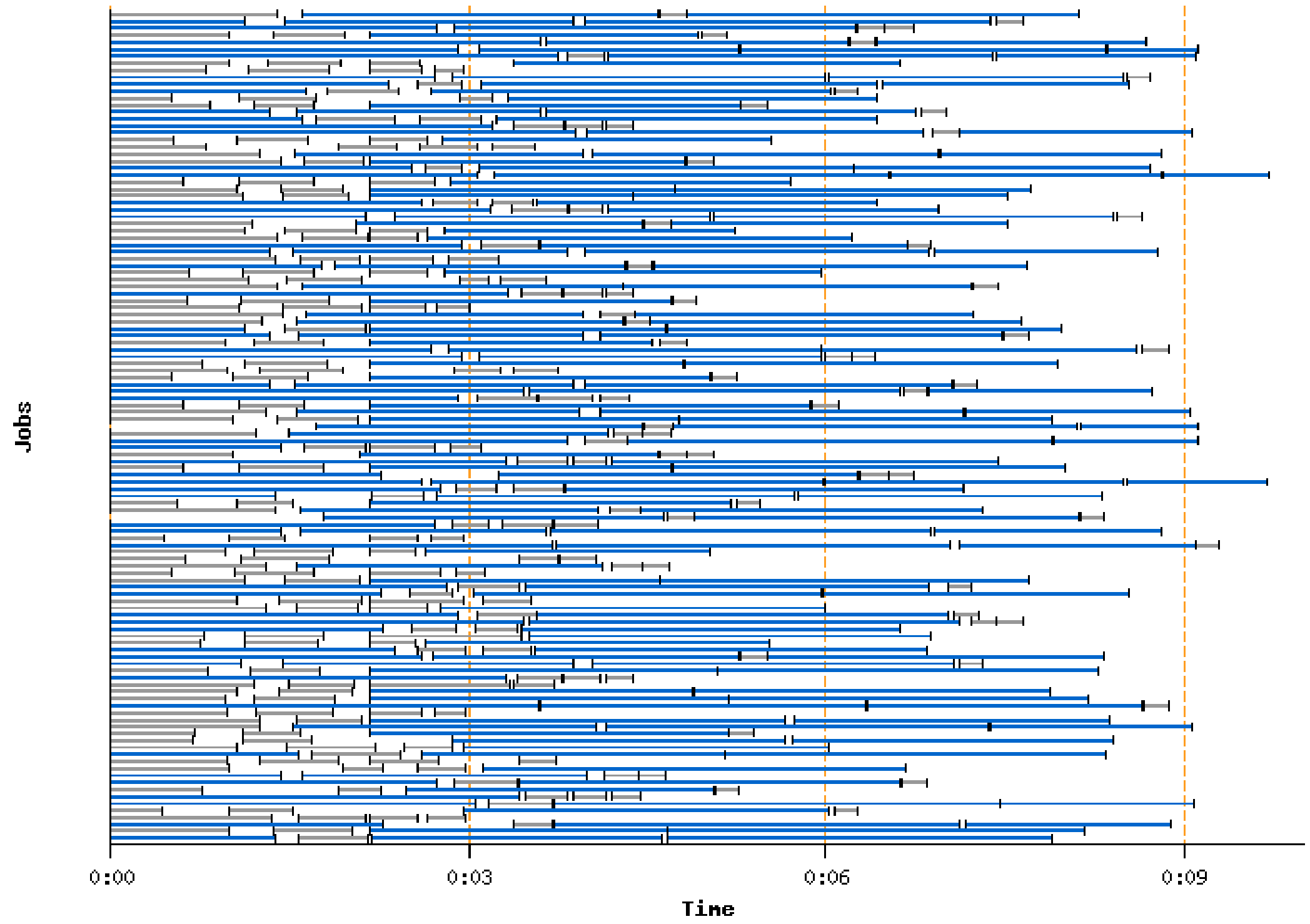

**Figure 11 Load Balancing Across Two Clusters**

We also notice the total execution time was cut down by 50% as compared with the execution time on a single cluster (ANL_TG). The reason was because Swift dispatched jobs to the two sites simultaneously, and the overall queue wait time was reduced significantly, as jobs were in two separate queues at the two sites, where each local scheduler dealt with only the jobs in its queue.

As illustrated in the former Section, Swift can achieve great speedup by submitting jobs to the Falkon service via the Falkon provider that we developed. Figure 12 shows both the throughput of Swift via the Falkon provider, and the throughput of the Falkon service (tasks were submitted from a Falkon client directly without going through Swift). In the experiment, the Falkon service was run on the ANL_TG site, and we submitted both from a separate node within that cluster (ANL->ANL), and from the UC_SUBMIT host (UC->ANL), and we measured the throughput, defined as the number of sleep(0) jobs completed per second. We ran two worker threads on 32 nodes (as each node has dual processors) to get the numbers for 64 nodes.

We measured a throughput of 120 tasks per second when submitting jobs from the Falkon client to the Falkon service directly in the same cluster, and a slightly lower throughput if we ran the Falkon client at the UC_SUBMIT host. When Swift was used to submit the same set of jobs, it was able to achieve up to 56 jobs per second in the LAN setting, and up to 46 jobs per second from the UC_SUBMIT host. Swift incurred a large overhead, but mainly because the scheduling in Swift involves creation of a sandbox environment for each job (so that it does not interfere with other concurrent running jobs on that node), and also logging and tracking of the job execution processing. For each sleep job, many extra steps such as site selection, execution directory setup and cleanup, exit code checking, etc were performed. In comparison, when jobs were submitted from the Falkon client, it involved only submission and getting back the exit code. Nevertheless, comparing with a standard setting when we use GT2 GRAM and PBS cluster scheduler where we can only get up to 2 jobs per second, the throughput of Swift combined with Falkon is improved by a factor of 23. These performance numbers with Falkon are with an older version that did not have all the latest optimizations we currently have in Falkon now; we expect these numbers to further increase with the latest Falkon code base that can achieve as high as 487 tasks/sec.

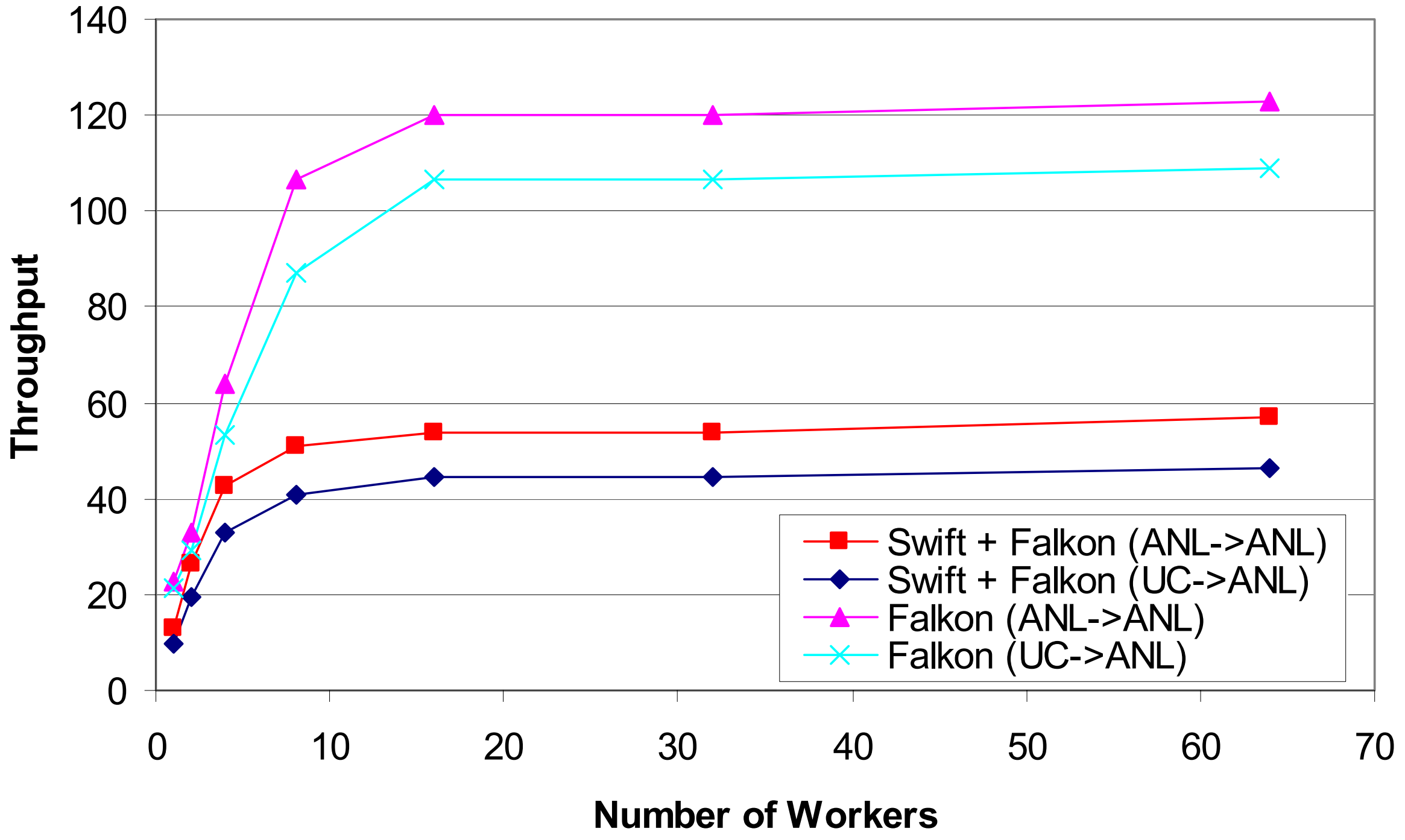

***Figure 12 Swift throughput with the Falkon provider***

We further compare the performance of submitting from Swift to providers such as PBS and Falkon using real application examples in the next Section.

## 5.4. Application Performance

Swift can either submit jobs via GRAM to a conventional batch scheduler such as PBS, in which case resource provisioning and job execution are combined together, or it can submit to the Falkon execution service, where resource provisioning is separated from job execution. We evaluate the effect of such separation using three examples: 1) fMRI (medical imaging), 2) Montage (astronomy), and 3) MolDyn (chemistry, molecular dynamics simulations). We repeat the fMRI and Montage experiments five times and the error bars indicate the standard deviations of the measurements. These applications all involve stages that have large number of small parallel jobs, a pattern that is common in many other scientific applications [41]. While we can push hundreds or even thousands of such small jobs via GRAM to a PBS queue, they will take up most of the available computing nodes but at the same time only achieve low throughput due to queuing overhead. We show that while clustering (bundling a number of small jobs into one large job) can help improve the throughput of such small jobs by 2 to 4 fold, the Falkon service can improve performance 7 to 10 fold for the fMRI application, in other words, achieving up to 90% reduction in execution time. For the Montage application, Falkon can achieve comparable performance to MPI execution. In the case of MolDyn, we achieved an 8 fold improvement in the application performance when comparing Falkon with GRAM/PBS submission.

### 5.4.1. fMRI (Medical Imaging Domain)

We note that for each volume, each individual task in the fMRI workflow required just a few seconds on an ANL_TG cluster node, so it is quite inefficient to schedule each job over GRAM and PBS, since the overhead of GRAM job submission and PBS resource allocation is large compared with the short execution time. In **Figure 13** we show the execution time for different input data sizes for the fMRI workflow.

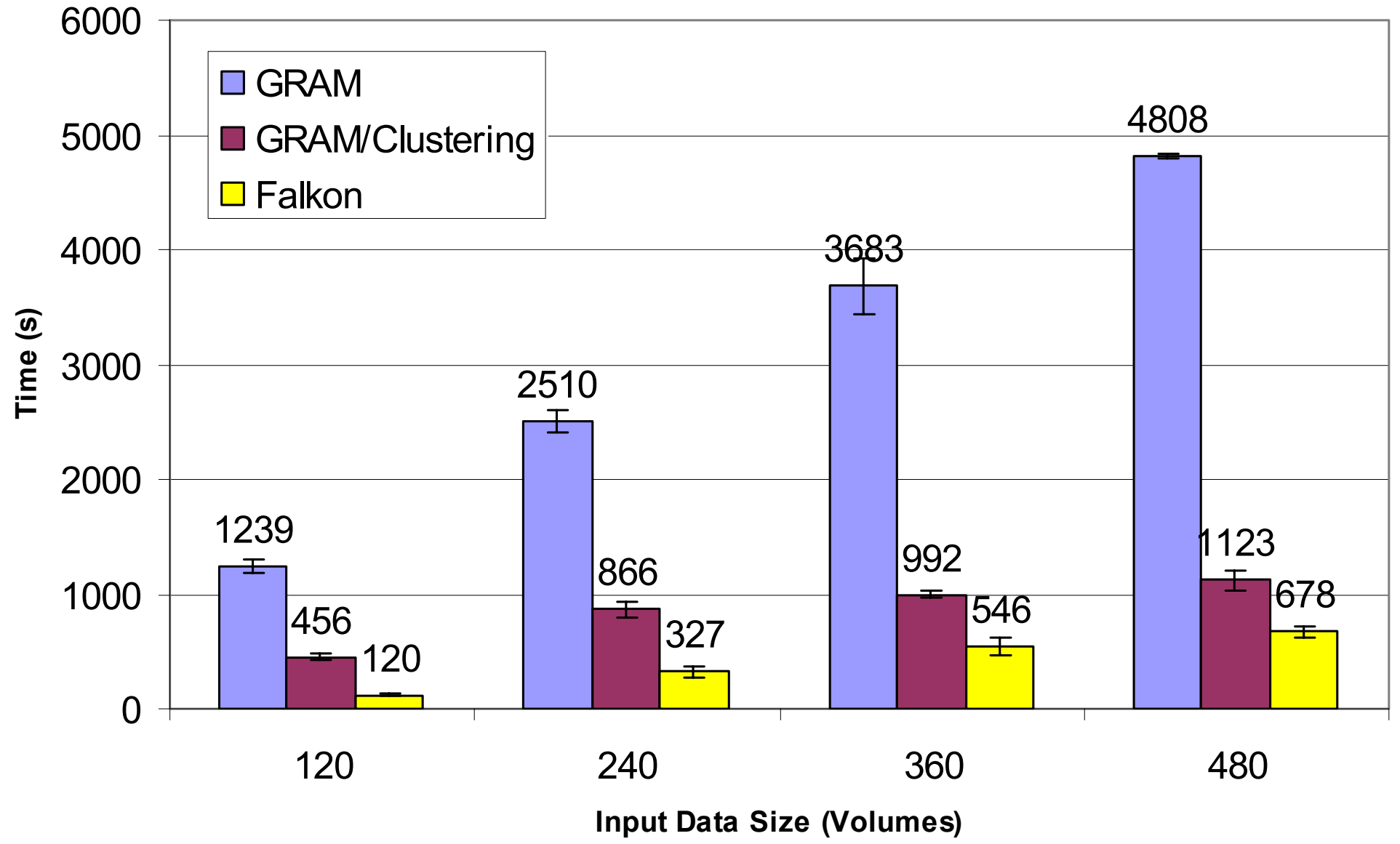

**Figure 13 Execution Time for the fMRI Workflow**

We submitted from UC_SUBMIT to ANL_TG and measured the turnaround time for the workflows. A 120-volume input (each volume consists of an image file of around 200KB and a header file of a few hundred bytes) involves 480 computations for the four stages, whereas the 480-volume input has 1960 computation tasks. The GRAM+PBS submission had low throughput although it could have potentially used all the available nodes on the site (62 nodes to be exact, as we only used the IA64 nodes). We can however bundle small jobs together using the clustering mechanism in Swift, and we show the execution time was reduced by up to 4 times (jobs were bundled into roughly 8 groups, as the grouping of jobs was a dynamic process) with GRAM and clustering, as the overhead was amortized by the bundled jobs. The Falkon execution service (with 8 worker nodes) however further cuts down the execution time by 40-70%, as each job was dispatched efficiently to the workers. We carefully chose the bundle size for the clustering so that the clustered jobs only required 8 nodes to execute. This choice allowed us to compare GRAM/Clustering against Falkon, which used 8 nodes, fairly. We also experimented with different bundle sizes for the 120-volume run, but the overall variations for groups of 4, 6 and 10 were not significant (within 10% of the total execution time for the 8 groups, plus or minus).

### 5.4.2. Montage (Astronomy Domain)

The Montage workflow demonstrated similar job execution time pattern as there were many small jobs involved. We show in Figure 14 the comparison of the workflow execution time using Swift with clustering over GRAM, Swift over Falkon, and MPI. The Montage application code we used for clustering and Falkon are the same. The code for the MPI runs is derived from the same set of source code, with the addition of data partitioning and inter-processor communication, so when multiple processors are allocated, each would process part of the input datasets, and combine the outputs if necessary. The MPI execution was well balanced across multiple processors, as the processing for each image was similar and the image sizes did not vary much. All three approaches needed to go over PBS to request for computation nodes, we used 16 nodes for Falkon and MPI, and also configured the clustering for GRAM to be around 16 groups.

The workflow had twelve stages, and we only show the parallel stages and the total execution time in the figure (the serial stages ran on a single node, and the difference of running them across the three approaches was small, so we only included them in the total time for comparison purposes). The workflow produced a 3x3 square degree mosaic around galaxy M16, where there were about 440 input images (2MB each), and 2,200 overlappings between them. There were two *mAdd* stages because we divided the region into subsets, co-added images in each subset, and then co-added the subsets together into a final mosaic. We can observe that the Falkon execution service performed close to the MPI execution, which indicated that jobs were dispatched efficiently to the 16 workers. The GRAM execution with clustering enabled still did not perform as well as the other two, mainly due to PBS queuing overhead. It is worth noting that the last stage *mAdd* was parallelized in the MPI version, but not for the version for GRAM or Falkon, and hence the big difference in execution time between Falkon and MPI, and the source of the major difference in the entire run between MPI and Falkon.

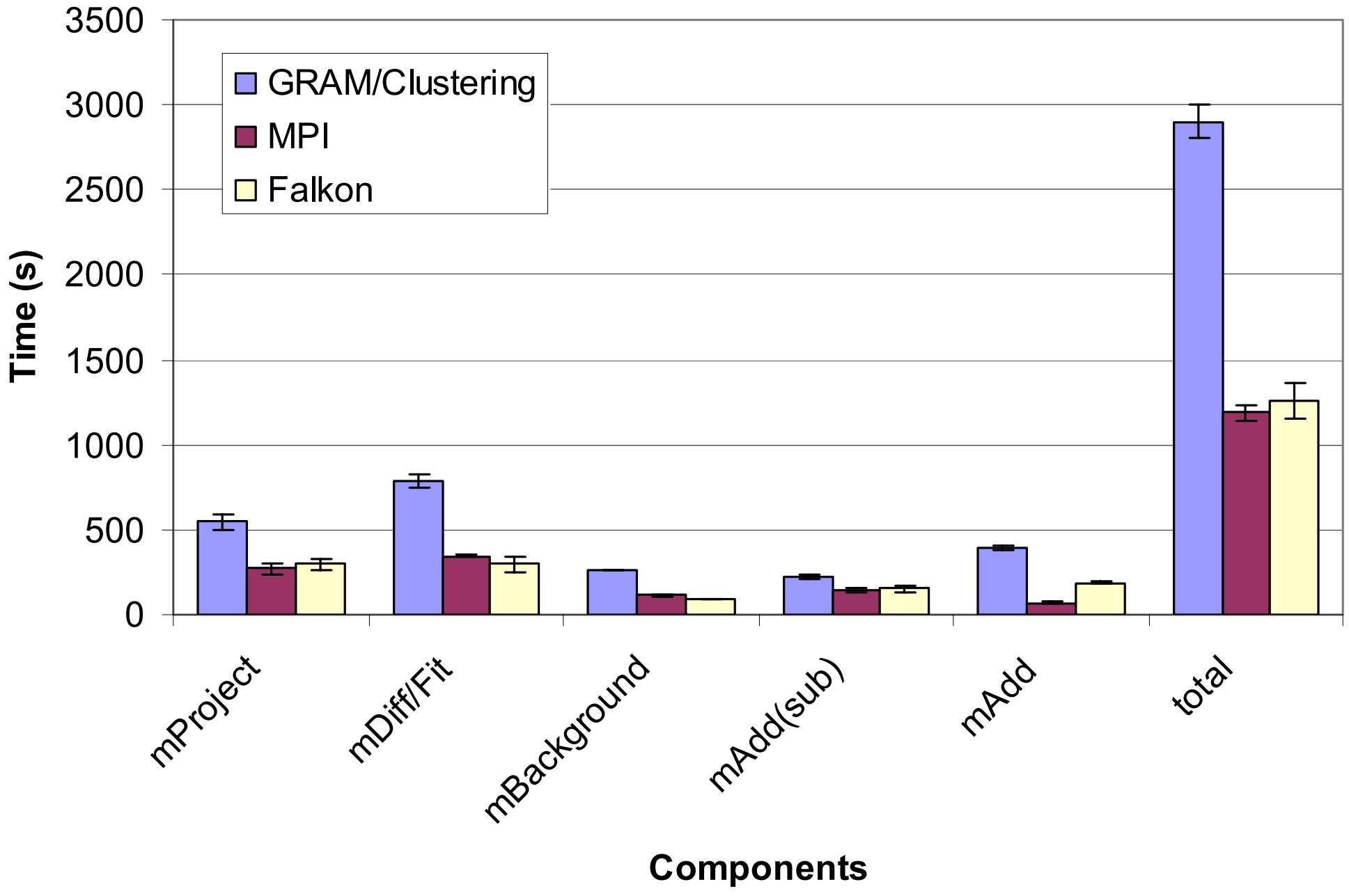

**Figure 14 Execution Time for the Montage Workflow**

Katz et al. [17] have also created a task-graph implementation of the Montage code, using Pegasus. They did not implement quite the same application as us: for example, they ran mOverlap and mImgtbl on the portal rather than on compute nodes, and they omitted the final *mAdd* phase. Thus direct comparison with Swift over Falkon is difficult. However, if we omit the final *mAdd* phase from the comparison, Swift over Falkon is then about 5% faster than MPI, and thus also faster than the Pegasus approach, as they claimed that MPI execution time was the lower bound for them. The reasons that Swift over Falkon performs better are that MPI incurs initialization and aggregation processes, which involve multi-processor communications, for each of the parallel stages, where Falkon acquires resource at one time and then the communications in dispatching tasks from the Falkon service to workers have been kept minimum (only 2 message exchanges for each job dispatch). The Pegasus approach used Condor's glide-in mechanism, where Condor is still a heavy-weight scheduler compared with Falkon.

### 5.4.3. MolDyn (Chemistry Domain)

We further illustrate the execution process in Falkon using a molecular dynamics (MolDyn) application. The goal of this study is to optimize and automate the computational workflow that can be used to generate the necessary parameters and other input files for calculating the solvation free energy of ligands, and can also be extended to protein-ligand binding energy. Solvation free energy is an important quantity in Computational Chemistry with a variety of applications, especially in drug discovery and design. The accurate prediction of solvation free energies of small molecules in water is still a largely unsolved problem, which is mainly due to the complex nature of the water-solute interactions. In the study, a library of 244 neutral ligands is chosen for free energy perturbation calculations. This library contains compounds with various chemical functional groups. Also, the absolute free energies of solvation for these compounds are known experimentally, and will serve as a tool to benchmark our calculations. All the structures were obtained from the NIST Chemistry WebBook database.

The MolDyn workflow consists of the following 8 stages:

- Stage 1: Annotate each of the molecules in the study with charges.
- Stage 2: For each molecule, the ligand structures are modified using Antechamber to generate the individual parameter and topology files for CHARMM Program. Antechamber automatically detects the atom type, bond type and bond order from the three-dimensional geometry of the molecule, and generates the residue topology file.
- Stage 3: The ligand structure is equilibrated using CHARMM program.
- Stage 4: Compute the solvation energy using three coupling (staging) parameters using the PERT module of the program CHARMM.
- Stage 6 computes the free energy for each of the resulting input configurations of each molecule in the study using weighted histogram analysis method (WHAM).
- Stages 7 and 8 extract the free energy value from prior stages and put the final data into a tabular form.

Stage 1 is done once for all molecules in the workflow, stages 2-8 are done for each molecule. The number of jobs in the workflow is about 1 + 84N where N is the number of molecules. The computation for each molecule takes about 235.4 minutes on one ANL_TG node processor, and 22.7 min given up to 32 ANL_TG nodes (64 processors); this resulted in a speedup of 10.4x, significantly less than 64 (the number of processors used at most), mostly due to the structure of the workflow which had several stages which were not parallelizable.  shows the graphical representation of each task for the 1 molecule experiment.  Red denotes wait queue time in Falkon and green denotes execution time per task.

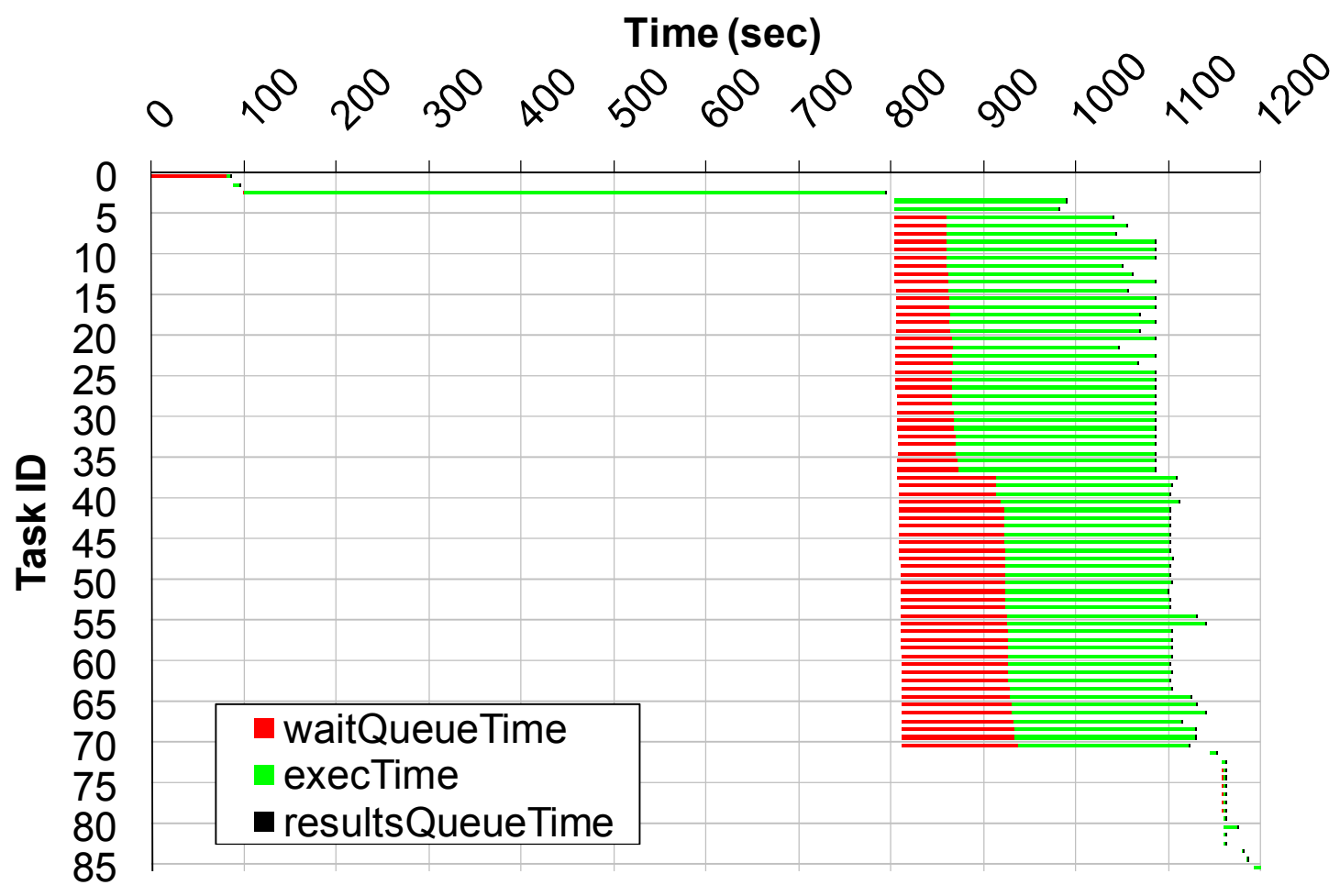

**Figure 15**: **MolDyn 1 Molecule Workflow Task View**

We were using the dynamic resource provisioning capabilities of Falkon in this experiment, and hence resources were only allocated on demand when they were needed. Notice the first job queue time is about 81 seconds, essentially the time it took from when Falkon requested one node and when it was allocated and ready to process work. Then, after the first three jobs completed in serial fashion), there was a stage with 68 independent jobs, which in turn triggered Falkon to allocated 31 more nodes (dual processors each) to be able to process all 68 independent jobs in parallel. The red lines for these 68 jobs shows the time delay that the allocation request took to traverse GRAM4 and PBS until the workers were ready to process jobs.

To better understand the structure of the MolDyn workflow, we show the graph structure of the workflow in **Figure 16**. Notice the large amount of parallelism that is obtainable at various stages of the workflow.

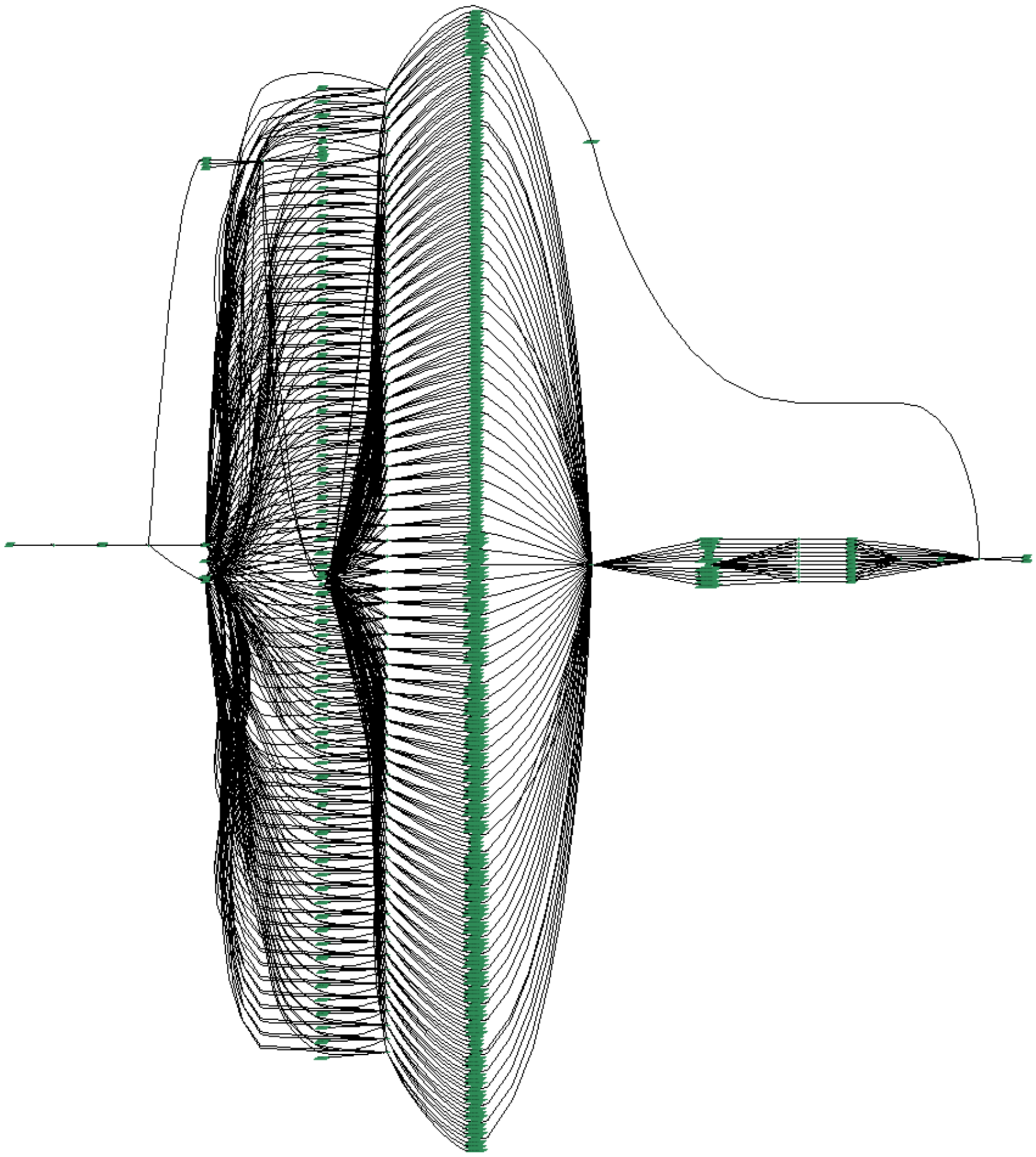

**Figure 16**: **MolDyn Workflow Graph**

showed the execution of 1 molecule, which is composed of 85 jobs that consume 235.4 CPU minutes. Our next experiment performed a 244 molecule run, which is composed of 20497 jobs that should take less than 957.3 CPU hours to complete; in practice, it takes even less as some job executions are shared between molecules. Figure 17 shows the resource utilization in relation to Falkon queue length as the experiment progressed. We see that as resources were acquired (using the dynamic resource provisioning, starting with 0 CPUs and ending with 216 CPUs at the peak), the CPU utilization was near perfect (green means utilized, red mean idle) with the exception of the end of the experiment as the last few jobs completed (the last 43 seconds). shows the same information on a per task basis, while Figure 18 shows the information on a per executor basis. The entire experiment with the exception of the last 43 seconds consumed 866.33 CPU hours and wasted 0.09 CPU hours (99.98971% efficiency); if we include the last 43 seconds as the experiment was winding down, the workflow consumed 867.1 CPU hours and it wasted 1.78 CPU hours, with a final efficiency of 99.7949013%. The experiment completed in 15091 seconds on a maximum of 216 processors, which results in a speedup of 206.9; note the average number of processors for the entire experiment was 207.26 CPUs, so the speedup of 206.9 reflects the 99.79% computed efficiency.

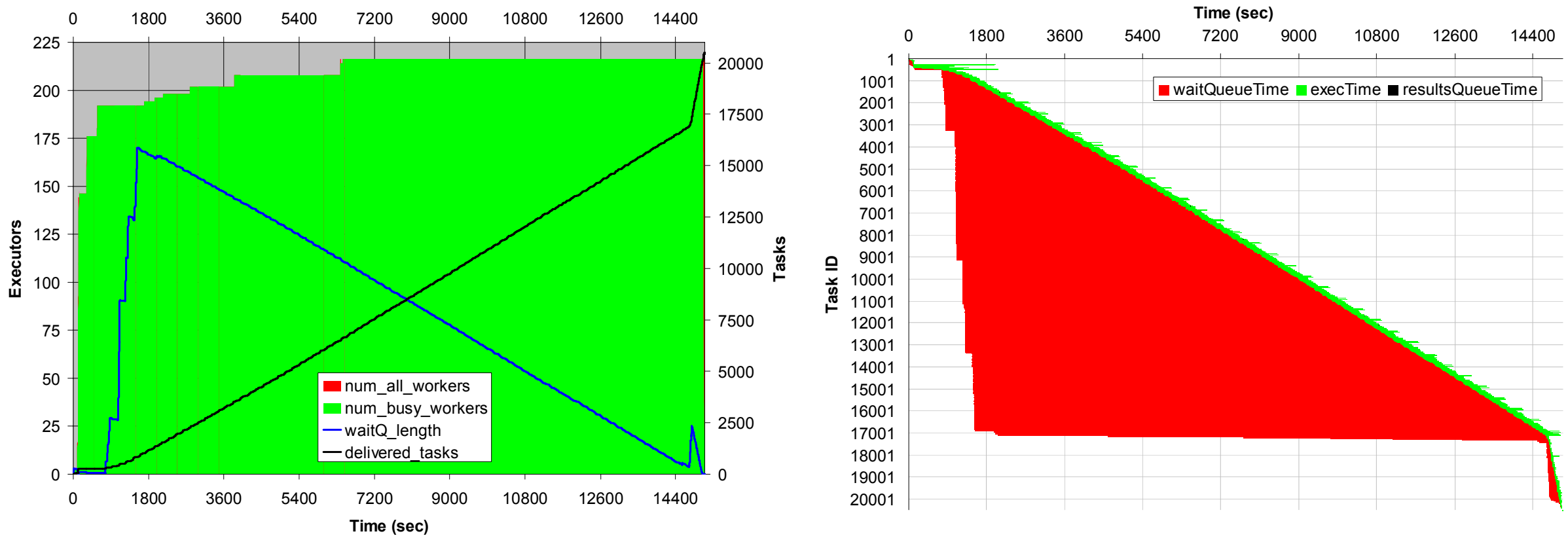

**Figure 17: 244 Molecule MolDyn application; left: summary view showing executor's utilization in relation to the Falkon wait queue length as experiment time progressed; right: task view showing per task wait queue time and execution time as experiment time progressed**

It is worth comparing the performance we obtained for MolDyn using Falkon with that of MolDyn over traditional GRAM/PBS. Due to reliability issues (with GRAM and PBS) when submitting 20K jobs over the course of hours, we were not able to successfully run the same 244 molecule run over GRAM/PBS. We therefore tried to do some smaller experiments, in the hopes that it would increase the probability of doing a successful run. We tried several runs with 50 molecules (4201 of jobs for the 50 molecule run, instead of 20497 jobs for the 244 molecule run); the best execution times we were able to achieve for the 50 molecule runs with GRAM/PBS (on the same testbed as we have used for Falkon, using up to 200 processors) took 25292 seconds. So we achieved a speedup of only 25.3X compared to 206.9X speedup when using Falkon on the same workflow and the same Grid site in a similar state.

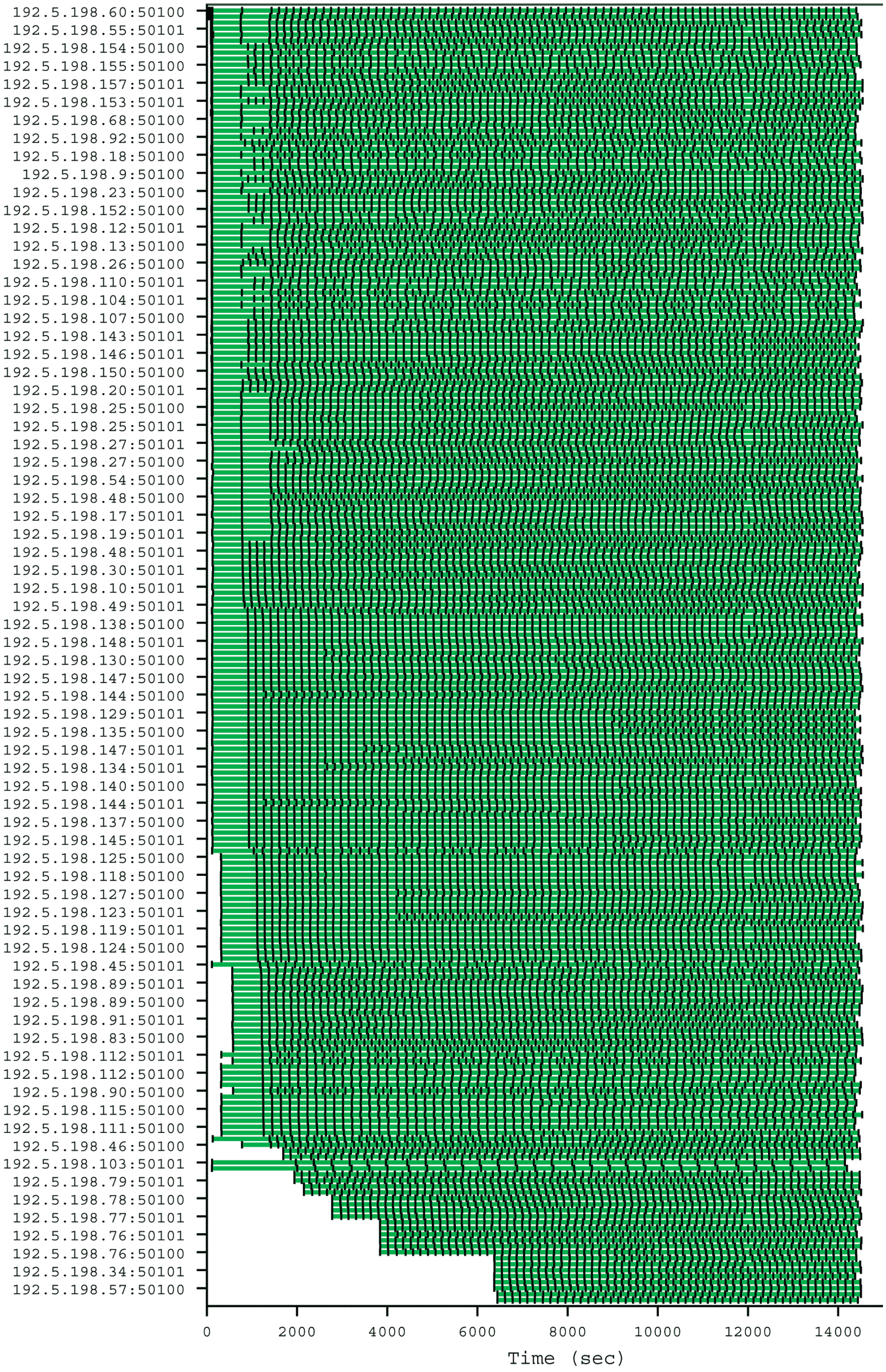

**Figure 18 244 Molecule MolDyn application: Executor's view showing each task that it executed (green) delineated by the black vertical bars showing each task boundary**

We explain this drastic difference mostly due to the typical job duration (~200 seconds) and the submission rate throttling of 1/5 jobs per second; the best case scenario is that the workflow could have kept 40 machines busy, but in reality the number of concurrent jobs ranged between 30~40 jobs. Increasing the submission rate throttle resulted in GRAM/PBS gateway instability, or even causing it to stop functioning. Furthermore, each node was only using a single processor of the dual processors available on the compute nodes due to the local site PBS policy that allocates each job an entire (dual processor) machine and does not allow other jobs to run on allocated machines; it is left up to the application to fully utilize the entire machine, through multi-threading, or by invoking several different jobs to run in parallel on the same machine. This is a great example of having a system like Falkon that allows the specific configuration of new queues that behave appropriately on a per application basis, which is quite impractical to do (mostly due to policy) in real-world deployed Grids.

### 5.4.4. Application Summary

The evaluation of the Swift runtime system shows that it is fast and scalable in executing large scale scientific computations. Swift leverages lightweight threading techniques and can schedule hundreds of thousands of parallel computations efficiently; its combination of type checking, retry mechanism and restart log supports reliable workflow execution. The abstract provider interface and various implementations allow workflows to be executed either on a single desktop, on a cluster managed by a batch scheduler, or on multi-site distributed resources. The integration with the lightweight Falkon execution service significantly improves system throughput for large number of small jobs. In particular, we show that Swift plus Falkon can achieve comparable performance to MPI execution for the Montage workflow, and also can reduce execution time by as much as 90% for the fMRI pipeline when compared with GRAM and PBS submission. We also showed a large scale MolDyn workflow with 20K tasks that was able to achieve 99.8% efficiency on 216 processors over a period of 15K seconds; Falkon was able to achieve an application speedup of 206.9X when GRAM/PBS was only able to achieve a 25.3X speedup on the same Grid site with similar usage conditions.

## 6. CONCLUSION

We have discussed the motivations for providing a high level abstraction of data, programs and computations in large scale Data Grid environments, and challenges inherent in attempting to do so, and we have proposed a scripting language for concise specifications of large scale parallel computations on file system data and a runtime system to support data, workflow, metadata, and provenance management. We have shown that the Swift language and system support efficient description, compilation, execution and optimization of parallel computations, using examples from large-scale scientific experiments in domains such as astronomy, biochemistry, and neuroscience.

SwiftScript is a parallel program notation that cleanly separates logical data structures from their physical storage formats, by means of the underlying XDTM mechanism. XDTM employs a two-level description of datasets, defining separately via a type system based on XML Schema the abstract structure of datasets, and the mapping of that abstract data structure to physical representations. The dataset mapping mechanism allows physical datasets to be mapped to logical data structures, thus enabling SwiftScript to represent dynamic workflow structures and expand workflow nodes during run time. SwiftScript models file system data as first-class typed objects, and allows for the definitions of concise typed procedures that operate on these typed datasets, and the composition of complex scientific workflows from simple and compound procedures. Its support for dataset selection and iteration enables a SwiftScript program to scale easily to tens of thousands of computations. Programs written in SwiftScript are much more compact in code size, and also easier to read, understand and maintain.

The runtime system Swift provides the automated mechanism to compile logical workflow representations into concrete distributed execution plan, manage and coordinate distributed computing, network and storage resource for the efficient and reliable execution of the plan, monitor and record the execution process, and trace the derivation history of data products. The provenance of data derivations is captured and stored in VDC along with dataset, procedure and workflow definitions; together they allow powerful exploration and expressive query of the virtual data products.

Swift integrates and extends the CoG Karajan engine, whose abstract provider interface supports submission to different computing facilities ranging from local desktops, to computer clusters and multi-site Grid environments. By leveraging the lightweight Falkon execution service, Swift is able to achieve high throughput in executing large number of fine-granular jobs, a common pattern we find in both the fMRI and Montage workflows, and other scientific applications.

In conclusion, the language and system together can bring significant usability and productivity increases, in representing and executing large scale scientific workflows, and in facilitating discovery, understanding, assessment, and sharing of data, procedures, workflows and computation resources, as well as in the productive use of distributed resources for computation, scheduling, and collaboration.

For future directions, we continue to work to improve the language and system to deal with other requirements and characteristics in large scale scientific computations. In particular, we are working on more intuitive invocation interface (the specification of how an application or service is invoked) in SwiftScript and easier mapper implementation. SwiftScript mappers are currently specified as Java classes. Pre-coded mappers work well for applications with structured file collections such as in the fMRI and Montage cases, but other customized mappers are required for more complex data structures, and the implementation of such mappers are not trivial.

Data partitioning is an important issue in harnessing parallel and distribute computing resources. Iterations in SwiftScript allow programs to operate on the finest level of datasets. However, SwiftScript currently does not support coarse level data partitioning, such as blocked or cyclic partition that are commonly observed in data parallel programs. We consider extending SwiftScript with (optional) data partitioning directives, and give experienced user more control over the data being processed.

We are also investigating data locality aware scheduling strategies. [43] Most cluster execution environments have a shared file system, where input and output datasets are staged in to and out from. This shared medium becomes a bottleneck when a large number of I/O intensive computations are executed in the cluster, although the computing resource may not have saturated. Considering that each node in a cluster usually has its own local disks, and local disk access is at least as fast, if not faster than a shared storage space, we can cache and replicate intermediate computation results on local disks, and make scheduling decisions according to the availability of the intermediate data, i.e. choosing the compute node that has most of the required data for a job. Such a strategy could further increase the number of parallel computations and improve system throughput.

Currently most of the applications we deal with are binary executables that operate on file system based data. However, our SwiftScript programming model can be equally applied to Web services invocation and coordination. SwiftScript specifies abstract interfaces to procedures, which can be bound to service invocations at runtime, and the XDTM data model is based on XML Schema, which can naturally capture Web service inputs and outputs. While Service Oriented Architecture is gaining adoption, SwiftScript can provide a bridge between the application and service deployment models.

# 7. ACKNOWLEDGEMENT

We would like to thank Bruce Berriman, Daniel Katz, and John Good for their help and support on the Montage application. Montage is supported by the NASA Earth Sciences Technology Office Computing Technologies program, under Cooperative Agreement Notice NCC 5-6261.

We would also like to thank Ben Clifford, Gregor Von Laszewski, and Tiberiu Stef-Praun for their contributions to the Swift system. The evaluation of the Swift system was carried out on computing facilities at the Computation Institute at the University of Chicago, and at the Teragrid ANL/UC site.

This work was supported by the National Science Foundation GriPhyN Project, grant ITR-800864 and iVDGL, grant PHY-122557; I2U2 – Interactions in Understanding the Universe, grant PHY-0636265; the Mathematical, Information, and Computational Sciences Division subprogram of the Office of Advanced Scientific Computing Research, U.S. Department of Energy under contract DE-AC02-06CH11357; the National Institutes of Health, grants NS37470, NS44393 and DC008638-01; and NASA Ames Research Center GSRP Grant Number NNA06CB89H.